\documentclass[prd,aps,showpacs,preprintnumbers,amsmath,amssymb,nofootinbib]{revtex4}
\usepackage[dvips]{graphicx}
\usepackage{epsf}
\usepackage{amsmath}
\usepackage{amssymb}
\usepackage{graphicx}% Include figure files
\usepackage{dcolumn}% Align table columns on decimal point
\usepackage{bm}% bold math
\def\be{\begin{equation}}
\def\ee{\end{equation}}
\def\bea{\begin{eqnarray}}
\def\eea{\end{eqnarray}}

\usepackage{color}

\newcommand{\comment}[1]{}

\def\EPSFIG[#1]#2#3#4{
\begin{figure}[H]
\begin{center}
\includegraphics[#1]{#2}
\end{center}
\caption{#3}
\label{#4}
\end{figure}
}

\begin{document}

%\date{\today}

\title{Defrosting in an Emergent Galileon Cosmology}

\author{Laurence Perreault Levasseur$^{a}$, Robert Brandenberger$^{a}$ and 
Anne-Christine Davis$^{b}$}

\affiliation{a) Department of Physics, McGill University,
Montr\'eal, QC, H3A 2T8, Canada}

\affiliation{b) Department of Applied Mathematics and Theoretical Physics,
Centre for Mathematical Sciences, Cambridge University, Cambridge, CB3 0WA,
United Kingdom}


\begin{abstract}

We study the transition from an Emergent Galileon condensate phase of the early
universe to a later expanding radiation phase. This ``defrosting" or ``preheating"
transition is a consequence of the excitation of matter fluctuations by the coherent
Galileon condensate, in analogy to how preheating in inflationary cosmology occurs
via the excitation of matter fluctuations through coupling of matter with the coherent
inflaton condensate. We show that the ``minimal" coupling of matter (modeled as a
massless scalar field) to the Galileon field introduced by Creminelli, Nicolis and
Trincherini in order to generate a scale-invariant spectrum of matter fluctuations is
sufficient to lead to efficient defrosting, provided that the effects of the non-vanishing
expansion rate of the universe are taken into account. If we neglect the effects of
expansion, an additional coupling of matter to the Galileon condensate is
required. We study the efficiency of the defrosting mechanism in both cases.

\end{abstract}

\maketitle

\newcommand{\eq}[2]{\begin{equation}\label{#1}{#2}\end{equation}}

\section{Introduction}
Over the past decades, more and more observational and theoretical discoveries and puzzles 
have hinted that maybe our current understanding of gravity does not encompass the whole 
picture. As such they motivated the study of modifications of the theory of general relativity, 
either through the introduction of new types of matter having unprecedented properties, or 
by modifying the way gravity itself propagates and couples to matter.

On one hand, the current observed accelerated expansion of the universe has caused 
a first clash with theoretical predictions \cite{Riess:1998cb, Perlmutter:1998np}. While the anthropic 
argument might be used to explain both the existence and the smallness of a cosmological 
constant \cite{Weinberg:1987dv}, the discovery of accelerated cosmological expansion has 
also engendered interest to explore theoretical possibilities for a modification of General Relativity (GR)
on cosmological scales, i.e. an infrared modification of GR. Along that line, there has been 
significant progress in developing screening mechanisms \cite{Khoury:2010xi}, such as the Chameleon mechanism \cite{2004PhRvD..69d4026K,
2004PhRvL..93q1104K,
2004PhRvD..70j4001G,
2006PhRvD..74j4024U,
Brax:2004px,
2004PhRvD..70l3518B}, the Symmetron mechanism 
\cite{2010PhRvL.104w1301H, 
2008PhRvD..77d3524O, 
2005PhRvD..72d3535P}, the Vainshtein mechanism 
\cite{Vainshtein:1972sx, 
2003AnPhy.305...96A, 
2002PhRvD..65d4026D} (which encompasses massive gravity theories 
\cite{2010PhRvD..82d4020D, 
2009PhLB..681...89G, 
2010PhLB..688..137D}, degravitation 
\cite{2002hep.th....9227A, 
2003PhRvD..67d4020D, 
2007PhRvD..76h4006D,
Patil}, brane induced gravity models 
\cite{Dvali:2000hr, Deffayet:2001pu,
2008PhRvL.100y1603D, 
2009PhRvL.103p1601D, 
2010PhRvD..81l4027D, 
2010PhRvD..81h4020A}, and Galileon theories \cite{Nicolis:2008in,2009PhRvD..79h4003D,
2009PhRvD..80f4015D,2010PhRvD..82f1501D, 2010JHEP...12..031P, Chow:2009fm, 
2011JHEP...01..099P, 2010PhRvD..82l4018H, 2011PhRvD..83d4042A, 
2011PhRvD..83h5015G, 2009PhRvD..80l1301S, Khoury:2011da, Trodden:2011xh}). 
These mechanisms are based on the assumption of extra scalar degree(s) of freedom, 
coupling gravitationally to both the baryonic and the dark sector, in such a way that the evolution 
of cosmological scales is affected to match the observed accelerated expansion, without 
being detectable through local experiments such as solar system tests of gravity
\footnote{Note that there are also suggestions that an instability to infrared (IR) fluctuations
might lead to a dynamical relaxation mechanism for the cosmological constant
\cite{Woodard,ABM,RHBrev2002}.}.

On the other hand, modified approaches to gravity might ameliorate some problems
which are encountered in the earliest stages of the evolution of the universe. The Null 
Energy Condition (NEC), which states that  $T_{\mu\nu}k^\mu k^\nu\geq 0$ for every null 
vector $k^{\mu}$, implies for a  Friedmann-Robertson-Walker (FRW) universe that $H$ 
always decreases, i.e. $\dot{H}\leq 0$. Going back in time, this leads to the initial cosmological
singularity. If the NEC is always satisfied, this might lead one to ask why is the universe expanding 
so rapidly in the first place. However, such a question is entangled with the question of the 
ultraviolet (UV) completion of gravity, because as one goes backward in time, $H$ and the energy
density increase, space contracts, until $H\sim m_{pl}$ and it becomes necessary to appeal to 
quantum gravity to understand the origin of the expansion of the universe. The objective of finding 
an alternative history of the universe in which quantum gravity effects do not become important 
has been a motivation to consider modifications of General Relativity by introducing new fields 
that would violate the NEC. On this basis many alternative cosmological scenarios have been proposed, such as string gas cosmology \cite{Brandenberger:1988aj, Nayeri:2005ck, Brandenberger:2006xi, Brandenberger:2006pr, Battefeld:2005av, Brandenberger:2008nx}, the
pre-Big-Bang scenario \cite{Gasperini:1992em, Gasperini:2002bn, Gasperini:2007vw}, 
models of Ekpyrotic, cyclic \cite{Khoury:2001wf, Khoury:2001bz,
Brandenberger:2001bs, Steinhardt:2001st, Finelli:2002we, Tsujikawa:2002qc, Tolley:2003nx,
Khoury:2003rt, Khoury:2004xi, Creminelli:2004jg} and bouncing cosmology (see
\cite{Novello} for a review of older work on bouncing cosmology and \cite{RHBrev2011}
for a review of more recent approaches), and even  higher dimensional inflation 
\cite{Steinhardt:2008nk}. 

A general class of such a violation of the NEC was studied formally 
for the first time in the context of ghost condensation \cite{ArkaniHamed:2003uy}. Instead of 
being doomed to contain disastrous instabilities as they were initially thought to 
\cite{Dubovsky:2005xd}, NEC-violating ghost condensates provided a general stable 
framework and opened many avenues for novel cosmological scenarios 
\cite{ArkaniHamed:2003uz, Creminelli:2006xe, Buchbinder:2007ad, Lin:2010pf}.

The second class of such models are obtained by making use of Galileons. Initially introduced in 
the context of the DGP model \cite{Dvali:2000hr,Deffayet:2001pu} and later proposed formally 
as a generic local infrared modification of General Relativity in order to explain the late-time
accelerated expansion, Galileons are a defined by introducing an extra scalar degree of 
freedom, $\pi$, kinetically mixed with GR \cite{Nicolis:2008in}. Demanding that they must obey 
the {\it Galilean symmetry},
\be
	\pi(x) \, \rightarrow \, \pi(x) + c + b_\mu x^\mu,
\ee 
imposing that this symmetry  also be a symmetry of the Langrangian, and that the $\pi$ equation 
of motion (EoM) be exactly second-order in derivatives leads one to the conclusion that in four 
dimensions, only five possible interaction terms are allowed in the Lagrangian, one per order 
of interaction. The fact that the EoM for $\pi$ is still only second order (regardless of the presence 
of higher order interaction terms in the Lagrangian) ensures that no ghost degree of freedom 
will stain the theory. Under these assumptions, Galileons provide a natural realization of the 
Vainshtein screening mechanism, with a self-accelerating solution at cosmological scales 
that decouples from short scales. 

However, Galileons also give rise to interesting cosmological scenarios when promoting 
the symmetry group of the Galileon Lagrangian from the Galilean symmetry to the conformal 
group SO(4,2). Among the three possible maximally symmetric solutions for $\pi$, choosing the 
time dependent one that breaks SO(4,2) to the isometry group of four-dimensional de Sitter 
space SO(4,1), one obtains a stable strong violation of the NEC in which the universe is 
expanding at late times, even in the case when it is not initially doing so. This scenario, dubbed 
the \textquotedblleft Galilean Genesis" \cite{CNT}, renders the usual assumption 
of an initial large and positive $H$ completely unnecessary.

More precisely, the \textquotedblleft Galileon Genesis" scenario postulates a universe that is asymptotically Minkowski ($\mathcal{M}_4$) in the past
\footnote{In this aspect the Galileon Genesis scenario is
a realization of the ``Emergent Universe Scenario" of \cite{Ellis}.}. 
Choosing the de Sitter solution $\pi_{dS}$
of the Galileon (which has zero energy density) in that limit will cause the scale factor to grow
exponentially as we approach $t\rightarrow0^-$. Perturbations around this background were 
previously shown to be stable, but they do not give rise to significant cosmological perturbations
and do not produce significant squeezing on large scales. The necessity for any cosmological 
scenario to produce the observed spectrum of scale-invariant primordial cosmological 
perturbations makes the introduction of a second scalar matter field, $\sigma$, necessary. 
Conformal symmetry requires that any other field coupling to the Galileon does so treating 
the Galileon as a dilaton, i.e. through an effective metric
\be \label{fictmetric}
	\mathfrak{g}^{f}_{\mu\nu} \, = \, e^{2\pi}g_{\mu\nu}.
\ee
If $g_{\mu\nu}$ can be approximated as $\eta_{\mu\nu}$, as is the case here asymptotically in 
the past, then $\sigma$ will behave as in a \textquotedblleft fake" de Sitter space and its dynamics 
will be the same as if space was undergoing inflation. In particular, the spectrum of perturbations in
the matter field will undergo squeezing and will be scale invariant. In the original paper proposing
this model \cite{CNT}, it was assumed that $\pi$ departs from its de Sitter solution 
(due to its coupling to gravity) and as energy is transferred to the matter field, $\rho_{tot}$ increases
until the system exits the regime of validity of the Galileon effective field theory. At that point, 
$\bigl( \mathcal{M}_4, \pi_{dS} \bigr)$  
ceases to be a valid background to expand around, and it was assumed that the energy density 
would be transferred to the standard adiabatic mode of regular matter, so that the universe would
then proceed to a standard radiation-dominated FRW phase, through a sort of 
\textquotedblleft defrosting" similar to reheating in standard inflation.

However, even if Galileon Genesis provides a successful implementation of an effective 
inflationary phase, the importance of the defrosting stage should not be underemphasized. It might 
be the case that the qualitative argument presented above conceals a graceful exit problem, 
for example if the Galileon fails to transfer sufficient energy density to the matter field  $\sigma$ 
to allow for a transition toward a radiation-dominated epoch. Hence one must ensure that 
a preheating stage transfers most of the energy density to the regular matter field.  Another 
source of concern is how the NEC-violating Galileon will react to a coupling with standard matter. 
It might be the case that, due to non-conventional kinetic properties, a rapid increase of the 
energy density in $\sigma$ will back-react on the Galileon by {\it accelerating} it instead of slowing 
it down as one would intuitively expect. In that case, again, the standard matter field will never 
come to dominate the evolution of $H$, and an evolution toward a radiation-dominated phase 
will not be possible.

Another point that makes the study of preheating/defrosting crucial is that it is this process 
that determines the amplitude of the adiabatic primordial cosmological perturbations produced. 
A viable cosmological model must produce an amplitude compatible with the 
$\delta\rho/\rho\sim 5\times 10^{-5}$ COBE normalization, and the amount of fine-tuming 
required to attain such an amplitude (if possible) in a given model gives information about 
the naturalness of the model. Moreover, the precise preheating mechanism describes how 
isocurvature perturbations, if they are produced either before of during reheating, will influence 
the observable adiabatic spectrum of perturbations.

In the case at hand, the fundamental growing fluctuations are produced in the matter field, 
while no sizeable perturbations are produced in the $\pi$ field. Since at the classical level 
in the background, the Galileon represents the adiabatic field, all sizeable scale-invariant perturbations are produced in the form of {\it entropy} fluctuations. In order to produce the 
observed adiabatic spectrum, these entropy modes must therefore be transferred to an 
additional adiabatic degree of freedom. However, the precise calculation of the amplitude 
of the $\delta\rho/\rho$ spectrum will not be the main focus of the current paper.  This is because 
this calculation is made more complicated by the absence of non-vanishing classical 
background, which renders $\delta\rho_\sigma/\rho_\sigma$ second order in the perturbations 
of $\sigma$ and which makes the spectrum of energy density perturbations in $\sigma$ 
qualitatively very different from the spectrum of field perturbations in $\sigma$ \cite{Durrer:1994}.

In this paper, we make the idea of preheating and defrosting more precise in the context of 
Galileon Genesis and ensure that the model does not suffer from a graceful exit problem. In 
Section \ref{ReviewofGG}, we review the Galileon Genesis formalism, first in the case of 
decoupling from gravity, where $\pi$ is identically in the de Sitter configuration, and then 
when coupling to gravity is re-established, in which case the de Sitter configuration in 
Minkowski space is only the limiting solution as $t\rightarrow-\infty$. In this more realistic 
model, $H\sim -1/t^{3}$ and the energy density in $\pi$ is created suddenly as $t$ approaches 
the singularity, which sets the right conditions for an efficient preheating. In Section 
\ref{preheatingSetup}, we review the preheating formalism. In Section \ref{PreAA}, we study 
how preheating proceeds in Galileon Genesis in the limit where gravity is decoupled from the
evolution of the fields. That is, we assume the de Sitter configuration for $\pi$ and $H=0$ as 
the background for the evolution of $\sigma$. It is found that, for minimal coupling between 
the Galileon and the matter field, the energy transfer to $\sigma$ is not sufficient to overcome the
growth of the Galileon as it evolves toward the singularity at $t=0$. To solve this problem, we 
explore the consequences of the introduction of further couplings between $\pi$ and $\sigma$, 
in the form of a potential term for $\sigma$. Such terms are chosen in such a way that they 
do not spoil the near scale-invariance of the spectrum of perturbations, but make it slightly 
red-tilted. Moreover, during the fictitious de Sitter phase, the amplitude of fluctuation modes 
remain time-independant after their freeze-out, which hints that this solution is as attractor, 
until the time of preheating, at which point they start growing. Upon the inclusion of such 
couplings, the energy transfer to $\sigma$ is made efficient enough for preheating to proceed.
Finally, it is found that in the case of both minimal and non-minimal coupling, the back-reaction 
of $\sigma$ on $\pi$ slows down the Galileon and making it evolve towards another 
maximally symmetric solution: $\pi=0$.

Finally, in Section \ref{Hneq0}, we reintroduce the coupling of the background to gravity. The
introduction of a growing $H$ and a Galileon departing from its de Sitter configuration has the
surprising effect of {\it accelerating} the growth of the energy density in the matter field close to the
singularity, making it fast enough to render obsolete the need for extra couplings between $\pi$ 
and $\sigma$. However, such an acceleration does not spoil the scale invariance of the spectrum 
at earlier times. The scale-invariant part of the spectrum remains slightly red-tilted, while a trough at scales corresponding to the scales freezing out at the beginning of preheating allows the UV end of the spectrum to be heavily blue-tilted. Hence the smallest scale modes freezing out during preheating dominate the energy density in $\sigma$ and permit efficient defrosting.  Moreover, the re-introduction of gravity does not spoil the back-reaction of $\sigma$, which is still found to slow down the Galileon.

% superluminality? should we talk about this?

\section{Review of Galileon Genesis}
\label{ReviewofGG}
\label{GalileonCosmoReview}

We will work with the simplest version of the Galileon minimally coupled to gravity.
In this case, the action of the Galileon scalar field $\pi$ is given by
\be \label{action}
S_{\pi} \, = \, \int d^4x \sqrt{-g} \left[ f^2 e^{2 \pi} (\partial \pi)^2
+ \frac{f^3}{\Lambda^3} (\partial \pi)^2 \box \pi + \frac{f^3}{2 \Lambda^3} (\partial \pi)^4 \right] \, ,
\ee
where $f$ sets the mass scale of the Galileon field (which is taken to be dimensionless)
and $\Lambda$ is a second mass scale which sets the energy at which the higher
derivative terms in the action become important. Lorentz indices are contracted with
the metric $g_{\mu \nu}$ and the ``box" operator  is built out of metric covariant
derivatives.

In the absence of coupling to gravity, there is a ``de-Sitter" solution $\pi_{dS}$ of the
equations which follow from (\ref{action}):
\be \label{sol}
e^{\pi_{dS}} \, = \, - \frac{1}{H_0 t}
\ee
valid in the time range $- \infty < t < 0$. In the above, the constant $H_0$ is given by
\be
H_0^2 \, = \, \frac{2 \Lambda^3}{3 f} \, .
\ee
As we shall discuss at the beginning of Section \ref{PreAA}, scalar matter fields
(like regular matter fields)
which are minimally coupled to the effective metric (\ref{fictmetric}) evolve as if
they were minimally coupled to a de Sitter metric.

The energy-momentum tensor of the Galileon field can be derived in the standard 
way and is given in \cite{CNT}. It can be verified that the solution (\ref{sol}) has 
vanishing energy density and pressure which scales as $- t^{-4}$. Thus, it has
an equation of state which violates the NEC. Since its pressure and the energy
density vanish as $t \rightarrow - \infty$ the solution (\ref{sol}) can be taken to
be the asymptotic solution in the far past even in the presence of gravity. Thus,
it corresponds to an emergent Universe which approaches Minkowski space-time
as $t \rightarrow - \infty$. The NEC violation allows for a transition to an
expanding phase. In fact, solving the Friedmann equations to leading
order in Newton's constant $G$ yields a background solution $\pi_0$
which scales as
\be
\pi_0 \, = \, \pi_{dS} -\frac{1}{2}\frac{f^2}{m_{pl}^2}\frac{1}{H_0^2t^2}
	\qquad  t\rightarrow-\infty \, 
\ee
with an associated Hubble constant which increases as
\be
H \, \approx \, -\frac{1}{3}\frac{f^2}{m^2_{pl}}\frac{1}{H_0^2t^3} \, .
\label{piwGrav-tinfinity}
\ee

The Hubble constant and the correction term in $\pi_0$ compared to the
de Sitter solution $\pi_{dS}$ increase without bound as $t \rightarrow 0$.
Hence the perturbative expansion in $G$ will break down at some $t\sim-H_0^{-1}f/m_{pl}$.
The equations in fact lead to a divergence in $H$ at some time $t_0$.
As shown in \cite{CNT}, the asymptotic behavior as $ t \rightarrow t_0$
is given by
\be
e^{\pi_0} \, \simeq \, \frac{8}{\sqrt{3}}\frac{f}{m_{pl}}\frac{1}{(t_0-t)^2} 
	\qquad t \rightarrow t_0 \, ,
\ee
with
\be	
H \, \simeq \, \frac{16}{3}\frac{f^2}{m_{pl}^2}\frac{1}{H_0^2(t_0-t)^3} \, .
\label{piwGrav-tclosesing}
\ee
The cosmological scale factor $a(t)$ then scales as
\be
a(t) \, \sim \, exp \left[ \frac{8 f^2}{3 H_0^2 M_{pl}^2} \frac{1}{(t_0 - t)^2} \right] \, .
\ee

The above solution describes a universe which emerges from a flat Minkowski
gravitational vacuum in the limit $t \rightarrow - \infty$ and then begins to expand
more and more rapidly (which is possible because the Galileon violates the NEC).
Eventually, $\pi$ becomes strongly coupled and the effective field theory
description of the Galileon breaks down. As the Galileon field grows in strength,
its coupling to regular matter fields becomes important, and it is to the study
of the effects of these couplings which we now turn.
%Should we add just a word about the stability of the background? 
 
\section{Preheating: Setup and Basic Equations}
\label{preheatingSetup}

If the Galileon genesis scenario is to successfully connect to late-time cosmology,
there needs to be a mechanism which drains energy-momentum from the Galileon
field and creates regular matter. This challenge is analogous to that faced in
inflationary universe cosmology. In an inflationary model \cite{Guth}, the energy density at
the end of the period of inflation is contained in the spatially homogeneous condensate of
the inflaton field, the scalar field responsible for generating inflation - in the same
way that at the end of the Galileon genesis phase the stress-energy is contained in
the spatially homogeneous Galileon field condensate. In the same way that couplings
between the inflaton field and regular matter need to be introduced to describe the
energy transfer at the end of inflation - a process called ``reheating" - coupling terms 
between the Galileon field and regular matter need to be introduced in Galileon 
cosmology. As is usually done in studies of reheating in inflationary cosmology (see
\cite{Reheatrev} for a recent review), we will model regular matter as another scalar
field. 

In the case of inflationary cosmology, reheating was first studied perturbatively 
\cite{Dolgov, AFW}. However, it was realized \cite{TB} that the perturbative analysis
misses out on the coherent nature of the inflaton condensate and in fact gives
completely wrong results for the duration of time the energy transfer takes.
It was shown \cite{TB} (see also \cite{DK}) that parametric resonance effects during
the oscillation of the inflaton condensate lead to a rapid energy transfer and
produce an out-of-equilibrium state of matter particles. This initial phase of
energy transfer was later \cite{KLS1} denoted ``preheating". The process in an
expanding cosmological background was then studied in more detail in \cite{STB, KLS2}.

As in the case of inflationary preheating, we expect coherence effects of the Galileon
condensate to be crucial when studying the energy transfer between the Galileon
background and matter, a process which we will call ``defrosting" of the Galileon
condensate state. Hence, we will employ the same formalism as is used in inflationary
preheating, namely a semiclassical analysis in which the linear matter field
fluctuations are quantized in the classical background given by the Galileon
condensate.  

Let us denote the scalar field representing matter by $\chi$ (in the
application to the Galileon genesis scenario this field will be the
$\sigma$ field mentioned earlier). We will treat
$\chi$ as a free scalar field. The non-trivial dynamics comes from the coupling
of $\chi$ to gravity and (in our case) to the Galileon. The first step in the
semi-classical analysis is to determine the canonically normalized matter field
$\tilde{\chi}$. For a standard kinetic term of $\chi$ (i.e. in particular in the absence
of coupling of $\chi$ to the Galileon), and in the case of minimal coupling of $\chi$
to gravity, the canonical field is
\be
\tilde{\chi} \, = \, a \chi \, .
\ee
The action then takes canonical form if we use conformal time $\tau$ related to
the physical time $t$ via 
\be
dt \, = \, a(t) d \tau \, .
\ee

We expand $\chi$ in terms of creation and annihilation operators $\hat{a_k}$ and
$\hat{a_k}^{\dagger}$ as
\be
\tilde{\chi}({\bf{x}}, t) \, = \, \frac{V^{1/2}}{(2 \pi)^{3}} 
\int d^3k \bigl( \tilde{\chi}_k(t)^{\star} \hat{a_k} e^{i {\bf{x}} {\bf{k}}}
+ \chi_k(t) \hat{a_k}^{\dagger} e^{- i {\bf{x}} {\bf{k}}} \, ,
\ee
where $V$ is the spatial cutoff volume and ${\bf{k}}$ is the comoving momentum vector.
The creation and annihilation operators obey the usual canonical commutation
relations. In the case of a free scalar field $\chi$ minimally coupled to gravity
and not coupled to the Galileon, the mode functions ${\tilde{\chi}}_k$ satisfy the equation
\be
{\tilde{\chi}}_k^{\prime \prime} + \omega_k^2 {\tilde{\chi}}_k \, = \, 0 \, ,
\ee
with
\be \label{freq}
\omega_k^2 \, = \, k^2 + m_{\chi}^2 a^2 - \frac{a^{\prime \prime}}{a} \,
\ee
where a prime indicates the derivative with respect to $\tau$ and $m_{\chi}$ is
the mass of $\chi$. 

Note that the effective square frequency $\omega_k^2$
can be negative if $a^{\prime \prime}/a$ is positive and if $k$ is sufficiently
small. This is the case for fluctuations with wavelengths larger than
the Hubble radius. On these long wavelength scales the fluctuation
amplitude increases while the microphysical oscillations freeze out.
This is the squeezing of fluctuations on super-Hubble scales
which is responsible for the growth and classicalization of quantum
vacuum perturbations in inflationary cosmology (see \cite{MFB, RHBrev2004}
for reviews of the theory of cosmological fluctuations and
\cite{Martineau, Kiefer} specifically for the question of classicalization) 
\footnote{Gravitational waves in an expanding universe undergo a similar
squeezing process \cite{Grishchuk}.}.
We will see in the next section that due to the coupling with the
Galileon condensate field, matter perturbations evolve as regular matter fluctuations
would in an effective time-dependent metric given by (\ref{fictmetric}).
Thus, long wavelength fluctuations are excited, leading to a
transfer of pressure from the Galileon to regular matter. Note
that in inflationary cosmology it is the coupling in the interaction
potential between the inflaton field and the matter field which leads
to the parametric excitation of long wavelength matter fluctuations.

In the semi-classical analysis we will assume that the ${\tilde{\chi}}$ field
starts out (mode by mode) in its vacuum state. 
With the field normalization chosen, taking expectation values of
$\chi$ correlation functions in an initial vacuum state corresponds
to calculating classical averages of these correlation functions using
as initial values of ${\tilde{\chi}}_k$ their harmonic oscillator ground
state values
\be
{\tilde{\chi}}_k (t_i) \, = \, \frac{1}{\sqrt{2k}} \, .
\ee

In the following, we will show that the same scalar field $\sigma$ which
was introduced in \cite{CNT} with applications for generating cosmological
perturbations in mind can provide a good model for the matter into
which the initial Galileon stress-energy flows. 

We briefly recall why the field $\sigma$ was introduced in \cite{CNT}.
The starting point is the observation that the initial spectrum of curvature
fluctuations induced by the Galileon field is blue if it stems from initial
vacuum fluctuations (which is the obvious initial state for fluctuations
in the emergent Galileon cosmology). It is a vacuum spectrum
with a spectral index $n_s = 3$ (scale-invariance corresponds to $n_s = 1$).
Since the curvature fluctuations are constant on super-Hubble scales,
its spectrum remains blue. Hence, a different mechanism is required
to generate a scale-invariant spectrum. In \cite{CNT} it was pointed out
that a massless scalar field which couples to the Galileon only through the
kinetic part of the Lagrangian (minimally coupled to the effective
metric (\ref{fictmetric})) acquires a scale-invariant spectrum since the
matter field evolves as if it were in de Sitter space (we will review
this result in the following section). Regular matter (modeled as a
massless scalar field) must couple to the Galileon in exactly the same way
\footnote{Treating regular matter as a massless field is a good approximation
since the mass of Standard Model matter fields is many orders of
magnitude smaller than typical mass scales relevant in the very
early universe.}. In the following section we will show that the induced growth
of fluctuations is strong enough to efficiently drain energy-momentum
from the Galileon, thus leading to successful ``defrosting" of the
cosmological Galileon condensate.

\section{Preheating: Analytical Analysis}
\label{PreAA}

\subsection{MinimalCoupling}

We start by considering the minimal interaction of the Galileon with the massless
scalar matter field $\sigma$ representing regular matter. The part of the action involving
$\sigma$ is:
\be
\label{Actionminimalcoupling}
	\mathcal{S}_I \, = \, \int d^4x\sqrt{-g}\mathcal{L}_I(\pi, \sigma )=\int d^4x\sqrt{-g}\left(-e^{2\pi}\partial_{\mu}\sigma\partial^{\mu}\sigma\right) \, .
\ee
Here, $\sigma({\bf{x}}, t)$ has units of energy, and the minus sign is required for $\sigma$ to 
have positive energy density, and hence to behave like a regular matter field. We first study 
the case in which gravity is decoupled, so that the indices are contracted with the Minkowski 
metric, $\eta^{\mu\nu}$. Moreover, $\pi$ is chosen to start out at $t\rightarrow-\infty$ and to 
follow the de Sitter background solution, so that $\sigma$ behaves as if it was in a 
\textquotedblleft fake" de Sitter background. The coupling is minimal in the sense that 
conformal invariance requires any coupling of $\sigma$ with $\pi$ to be through the 
\textquotedblleft fake" metric $\mathfrak{g}^f_{\mu\nu}=e^{2\pi}\eta_{\mu\nu}$.

The equation of motion (EoM) for $\sigma$, upon the field rescaling 
\be \label{rescaling}
\sigma \, \rightarrow \, u^{-1}(t)\tilde\sigma \, \equiv \, e^{-\pi}\tilde{\sigma}
\ee
to obtain the canonically-normalized variable, and upon performing the Fourier transform 
\be
\tilde\sigma(\mathbf{x}, t) \, = \, \int \frac{d^3k}{(2\pi)^3}e^{i\mathbf{k}\mathbf{x}} \sigma_k(t) V^{1/2} \, ,
\ee
(where $V$ is the cutoff volume coming from putting the theory in a finite box), is then:
\be
\label{sigmaEoM}
	\ddot{\tilde\sigma}_k + \left(k^2-\frac{2}{t^2}\right)\tilde\sigma_k \, = \, 0 \, .
\ee
As discussed in Section \ref{preheatingSetup}, we
want to match this with the usual equation for the canonically normalized massless
scalar matter field $\chi$ in a fixed background $a(\tau)$:
\be
	{\chi}_k''+\left(k^2-\frac{a''}{a}\right)\chi_k \, = \, 0 \, ,
\ee
where the primes denote derivatives with respect to the conformal time $\tau$. 
In the case of a de Sitter background, we have
\be
\frac{a^{\prime \prime}}{a} \, = \, \frac{2}{\tau^2} \, .
\ee
Thus, the analogy 
between the two EoMs is clear. In our case, however, the real metric is flat Minkowski, which 
means $a  = 1$ and $\tau = t$. However, the function $a(t)$ from the inflationary case 
matches to an expression $\mathfrak{a}^{f}(t)$, the \textquotedblleft fake" scale factor, 
in (\ref{sigmaEoM}). This way, the standard analysis of inflationary cosmology can be applied to 
canonically quantize the matter field and get the power spectrum of perturbations in 
$\sigma_k$ at the time when they freeze out. 
Note however that, in contrast to the case of massless inflation, the field rescaling 
$u(t)$ used earlier  in (\ref{rescaling}) and $\mathfrak{a}^{f}(t)$ defined here need {\it not}  
be equal.  Even though they are for the simple coupling with $\pi$ considered here, as soon as a 
potential term for $\sigma$ is introduced (as will be done in the following
subsection), they cease to be.

For the de Sitter Galileon background $\pi_{dS}$, the fake scale factor satisfies the equation 
$\ddot{\mathfrak{a}}^f/\mathfrak{a}^f=2/t^2$. The solutions for $\mathfrak{a}^{f}$ are hence
\be
\mathfrak{a}^{f} \, = \, c_1 t^2 + \frac{c_2}{t} \, ,
\ee
where $c_1$ and $c_2$ are constant coefficients. As in our case $t \rightarrow 0$, the 
growing mode is selected and we can use $\mathfrak{a}^f=1/H_0t$. We choose 
$\mathfrak{a}^f$ to be unitless for simplicity, but note that the overall normalization and units of 
$\mathfrak{a}^f$ bear no physical significance, and are therefore irrelevant. 
All of the physical information is enclosed in the scaling $u(t)$, whose normalization is fixed by 
the requirement of transforming the kinetic term of $\sigma$ in (\ref{Actionminimalcoupling}) 
into a canonical kinetic term in flat space for $\tilde{\sigma}$.

Before turning to our analysis of defrosting of the Galileon background via production
of $\sigma$ excitations, we will review why a scale-invariant spectrum of fluctuations of
$\sigma$ emerges \cite{CNT}.
For a given $k$, a mode of $\tilde\sigma_k$ will oscillate with constant amplitude as 
long as the corresponding length scale stays within the fake Hubble radius, i.e. for
\be 
k^2 \, > \, k_{frz}^2 \, \equiv \, \ddot{\mathfrak{a}}^{f}/\mathfrak{a}^{f}(t_{frz}) \, = \, 2/t_{frz}^2 \, ,
\ee
and it will freeze out at $k = k_{frz}$. For $k < k_{frz}$, there is a mode of $\tilde\sigma_k$ which grows as $\mathfrak{a}^{f}$.  Hence, the spectrum of the field $\sigma$ will be given by:
\bea
	\langle \sigma_k\sigma_{k'}\rangle=(2\pi)^3\delta(\mathbf{k}+\mathbf{k}')\left|\sigma_k\right|^2&=&(2\pi)^3\delta(\mathbf{k}+\mathbf{k}')\left|u^{-1}(t)\tilde\sigma_k(t)\right|^2\nonumber \\
	&=&(2\pi)^3\delta(\mathbf{k}+\mathbf{k}')\left(u(t)\right)^{-2}\frac{\left(\mathfrak{a}^{f}(t_{frz})\right)^{-2}}{\left(\mathfrak{a}^{f}(t)\right)^{-2}}\left|\tilde\sigma_k(t_{frz})\right|^2 \nonumber \\
	&=&(2\pi)^3\delta(\mathbf{k}+\mathbf{k}')\left(u(t)\right)^{-2}\frac{\left(\mathfrak{a}^{f}(t_{frz})\right)^{-2}}{\left(\mathfrak{a}^{f}(t)\right)^{-2}}\left|\tilde\sigma_k(t_{i})\right|^2 \nonumber \\
	&=&(2\pi)^3\delta(\mathbf{k}+\mathbf{k}')(H_0t_{frz})^2\left|\frac{1}{\sqrt{2k}}\right|^2 \nonumber \\
	&=&(2\pi)^3\delta(\mathbf{k}+\mathbf{k}')\frac{H_0^2}{k^3} \, .
\eea
Hence the spectrum of perturbations is scale invariant. 

We are now interested to know, first, whether the energy density transferred to $\sigma$ from 
$\pi$ is sufficient for the matter field to overcome the Galileon energy-wise. Second, we want 
to ensure that this process of energy transfer will, as one would intuitively think, indeed 
back-react on $\pi$ to slow it down and make it evolve toward the $\pi=0$ Lorentz invariant 
vacuum solution. The fulfilment of these two conditions will ensure that the 
\textquotedblleft fake" de Sitter phase eventually comes to an end, and, as the $\pi$ field is 
driven to zero by the growth of $\sigma$, that it will be followed by a radiation-dominated epoch.

The energy density in $\sigma$ can be computed from the stress-energy tensor of the part
of the action involving $\sigma$, namely $\mathcal{S}_I$:
\be
\label{stress-energy}
	T_{\mu \nu}(\sigma) \, = \, \left[ 2\partial_\nu \sigma \partial_\nu \sigma - g_{\mu\nu} (\partial \sigma)^2\right] e^{2\pi} \, .
\ee
Note that this expression involves the $\sigma$ field in position space, not the rescaled 
field $\tilde{\sigma}$). The energy density and pressure in $\sigma$ for the chosen 
$\pi$ background are thus given by:
\bea
	\rho_\sigma(\mathbf{x},t) = \frac{1}{(H_0t)^2}\left( \dot{\sigma}^2+\frac{(\nabla\sigma)^2}{a^2}\right) &\Rightarrow\bar{\rho}_\sigma&=\frac{1}{V}\int d^3x\rho_\sigma(\mathbf{x},t) \nonumber \\
	&& = \int\frac{d^3k}{(2\pi)^3} \frac{1}{(tH_0)^2}\left( \dot{\sigma}_k^2+\frac{k^2\sigma_k^2}{a^2}\right)\nonumber \\
	&&\equiv\int\frac{d^3k}{(2\pi)^3}\rho_\sigma(\mathbf{k},t)
\eea
\bea
	p_\sigma(\mathbf{x},t) =  \frac{1}{(H_0t)^2}\left( \dot{\sigma}^2 -\frac{1}{3}\frac{(\nabla \sigma)^2}{a^2} \right) &\Rightarrow \bar{p}_\rho&=\int \frac{d^3k}{(2\pi)^3} \frac{1}{(tH_0)^2}\left( \dot{\sigma}_k^2 -\frac{1}{3}\frac{k^2 \sigma_k^2}{a^2} \right) \nonumber \\
	&&\equiv\int\frac{d^3k}{(2\pi)^3}p_\sigma(\mathbf{k},t) 
\eea
where by $\bar{\rho}_\sigma$ and $\bar{p}_\sigma$ we mean the space averages of these 
quantities over the spatial volume $V$, and we have then used the Fourier modes of $\sigma$ 
to re-write the expressions. We perform such a spatial average for the sake of comparison 
with $\rho_\pi$, which is a homogeneous quantity. 
Moreover, in the last step by defining $\rho_\sigma(\mathbf{k},t)$ and $p_\sigma(\mathbf{k},t)$ we mean the energy density and pressure in each $k$-modes contributing to the spatial averages $\bar{\rho}_\sigma$ and $\bar{p}_\sigma$ respectively, not the Fourier transform of $\rho_\sigma(\mathbf{x},t)$ and $p_\sigma(\mathbf{x},t)$. 

To know the amount of energy density transfered to the $\sigma$ field, we need to integrate 
the energy density over all $k$-modes. Since we started out in Minkowski space at $t=-\infty$, 
$H$ is initially zero, so $H_0$ and $k$ are physical (as opposed to comoving). Hence the scales 
of interest to us are the ones with $\lambda\sim1mm$, or $k_{i}=10^{-31} l_p^{-1}$, which 
give the wavelength of modes corresponding to the current large scale structure 
at the time of reheating (if the scale of
reheating is taken to be comparable to the scale of particle physics Grand Unification). We 
set the IR cutoff to be a bit larger than that scale.

Also, we note that modes below the scale $k_{frz}$ are still in their stage of quantum
vacuum oscillation, and hence they do not contribute  to the renormalized energy density
(obtained by subtracting the vacuum contribution). The UV cutoff in the integral over 
momenta at time $t$ is therefore set by the smallest scale for which the mode functions 
have frozen out at the time at the time $t$.  This value of $k$ is given by $k_{frz}$, and
it corresponds to the ``fake" Hubble radius at time $t$. 
%which is well bellow $k\sim\mathcal{O}(1)$. 
%Note that we only integrate up to this \textquotedblleft fake" Hubble radius scale, not up to 
% the actual Hubble radius as one normally does (one should recall that we assumed 
% decoupling from gravity, which implies $H=0$).  

We use the solution 
\be
\sigma_k(t) \, = \, u(t)^{-1}\frac{\left(\mathfrak{a}^{f}(t_{frz})\right)^{-1}}{\left(\mathfrak{a}^{f}(t)\right)^{-1}}\tilde\sigma_k(t_{i}) \, ,
\ee
from above to express the growing mode of $\sigma_k$ that will contribute the most to 
$\rho_\sigma(\mathbf{k},t)$. We can then express the energy density in each $k$-mode 
satisfying $k \ll k_{pre}$, that is, every mode that is frozen out, as:
\be
	\rho_\sigma(\mathbf{k}, t) \, = \, \frac{1}{k t^2} \, .
\ee
Hence
\be
\label{rhoSigdeGravMin}
	\bar\rho_\sigma \, = \, \int_{k_i}^{k_{frz}} \frac{d^3 k}{(2\pi)^3} \rho_\sigma(\mathbf{k}, t)
	\, = \, \frac{1}{2\pi^2t^2}\left( \frac{1}{t^2}-\frac{1}{t_i^2} \right)  \, ,
\ee
where we have used $t_i=\sqrt{2} / k_i$, and $k_{frz}=\sqrt{2}/t$. 
%% Hence, if we assume decoupling from gravity, a supplemental massless degree of freedom 
%% will naturally be given a scale invariant spectrum of perturbations. 
We see that energy density transfer from $\pi$ to the matter field $\sigma$ will only be 
sufficient for $\rho_\sigma$ to grow as $\sim1/t^4$ close to the time of 
defrosting/preheating as $t \rightarrow 0$. 

Note that the energy density in $\sigma$ scales with the same power of $t^{-1}$ 
as the background pressure of the Galileon condensate (which is 
Minkowski spacetime with $\pi = \pi_{dS}$ and $\rho_{\pi} = 0$, $\pi = \pi_{dS}$). Since the 
energy density of $\pi_{dS}$ vanishes, we see that $\rho_\sigma$ will immediately 
dominate the total energy density. Naively, this might lead us to expect that the 
universe should defrost/preheat quickly (we shall come back to this point soon). But
it also makes clear that the back-reaction of  $\sigma$ on $\pi$ could be very
important. Hence, we now turn to the analysis of this back-reaction, with the goal to 
ensure that it will slow down the Galileon from its de Sitter solution toward the solution $\pi=0$. 

Including the variational derivative of $\mathcal{S}_I$, the equation of motion for $\pi$ gives:
\bea
\label{firstbackreactpi}
	\ddot{\pi}\left[ 1 - \frac{2}{H_0^2}e^{-2\pi}\dot{\pi}^2 + \frac{4}{3H_0^2}e^{-2\pi}\nabla^2\pi\right] \, 
	&=& \, -\dot{\pi}^2  + (\nabla\pi)^2+\nabla^2\pi +
	        \frac{2}{H_0^2}e^{-2\pi}\nabla^2\pi(\nabla\pi)^2 \nonumber \\
	 & & \, + \frac{2}{3H_0^2} e^{-2\pi} \left[2(\nabla\dot\pi)^2 - \dot\pi^2(\nabla\pi)^2 -
	        \dot\pi\nabla\dot\pi\nabla\pi-\dot\pi^2\nabla^2\pi\right] \nonumber \\
	& & \, -\frac{1}{f^2}\left[\dot{\sigma}^2-(\nabla\sigma)^2 \right] \, .
\eea

We are interested in studying the back-reaction of the linear fluctuations of $\sigma$ (computed
above) on the background of $\pi$. We use the following expansion of $\pi$ in a fixed 
Minkowski background:
\bea
	\pi \, &=& \, \pi_{dS} + \delta\pi + \delta^{(2)}\pi \\
	\sigma \, &=& \, \delta\sigma + \delta^{(2)}\sigma\, ,
\eea
where the background is homogeneous, but higher order perturbations are allowed not to be. The
expansion parameter is the amplitude of the linear fluctuations. The term $\delta \sigma$
is the linear fluctuations in $\sigma$ which we have just studied. The term $\delta \pi$
corresponds to the linear fluctuations in $\pi$. However, we already know from section 
\ref{GalileonCosmoReview} that inhomogenous adiabatic linear perturbations in $\pi$ will be
cosmologically irrelevant. 
%so that effectively there can be no such sizeable perturbations. 
Hence, when expanding (\ref{firstbackreactpi}) to linear order, we will find no significant growing 
solutions because all contributions from $\sigma$ will be second order and $\pi_{dS}$ is an 
attractor. We can directly go to second order and neglect the contributions from $\delta\pi$. We obtain
\be
	 \delta^{(2)}\ddot\pi - \frac{2}{t}\delta^{(2)}\dot{\pi} - \frac{4}{t^2}\delta^{(2)}\pi-\frac{1}{3}\nabla^2(\delta^{(2)}\pi) \, = \, \frac{1}{f^2}\left[\delta\dot\sigma^2-(\nabla\delta\sigma)^2\right]  \, .\label{homobackreact2}
\ee
In order to have a better intuition of the way the $\sigma$ source terms will drive the 
Galileon backreaction, we go to Fourier space. We are mainly interested in the 
homogeneous back-reaction $\delta^{(2)}\pi_{k=0}$. In the equation for that zero mode, 
the $\sigma$ terms will turn into an integral over $k$-space where each $\sigma_k$ mode 
couples with the corresponding $\sigma_{-k}$ mode. To perform that integral we use, as before, 
the rescaled field $\tilde\sigma_k$ and the fake scale factor $\mathfrak{a}^f$ to express 
$\sigma_k$ for $k < k_{frz}$ as $\sigma_k = H_0/k^{3/2}$. Finally, the modes $k   > k_{frz}$ are 
oscillating with constant amplitude, so that they do not contribute to the renormalized value of
the integral. Once the integral is performed, we are left with the following equation for the 
back-reaction on the Galileon background:
\be
	 \delta^{(2)}\ddot\pi_{k=0} - \frac{2}{t}\delta^{(2)}\dot{\pi}_{k=0} - \frac{4}{t^2}\delta^{(2)}\pi_{k=0}
	 \, = \, -\frac{H_0^2}{2\pi^2f^2} \left(\frac{1}{t^2}-\frac{1}{t_i^2} \right) \, .
\ee
Since the source term is negative (recall that $t > t_i$), we thus conclude that the growth of 
$\sigma$ with time will indeed slow down the Galileon background, and this way will make it 
move from the de Sitter solution toward the $\pi=0$ solution. From this analysis, we can 
hope that the defrosting/preheating indeed proceeds and ends the fake de Sitter phase, 
leading to a radiation dominated expanding phase \footnote{Note, in particular, that there
is no instability in the system - one might have feared that the excitation of $\sigma$ would lead
to an increase in the amplitude of $\pi$.}.

However, if we now go back to the comparison of the growth of $\rho_\sigma$ relative to 
$\rho_\pi$, and, for the purpose of comparison, we look at a more realistic background in 
which the coupling to gravity has been re-introduced close to the time of preheating, we see that, 
due to the singularity in the solution, $\rho_\pi$ grows as $\sim1/t^6$ as $t\rightarrow0$. This 
means that the growth of $\rho_\sigma$ will {\it not} be fast enough to overcome  $\pi$ in a 
more realistic setup if we neglect the effects of a non-zero real $H$ on $\sigma$.

There are few avenues to overcome this apparent difficulty. A first one is to introduce further 
interaction terms in $\mathcal{S}_I$ that would make the coupling between $\pi$ and $\sigma$ 
stronger only close to the singularity. This way, the scale invariance of the spectrum of $\sigma$
perturbations would be preserved, but the growth of $\rho_\sigma$ close to the singularity could 
be made much faster. This is the avenue we explore next.

\subsection{Non-Minimal Coupling}

Our goal here is to add coupling terms between the Galileon and the matter field in $\mathcal{S}_I$ 
that make $\rho_\sigma$ grow faster than $\rho_\pi$ as $t\rightarrow0$, in such a way that the 
scale invariance of the spectrum of $\sigma$ perturbations is preserved. That is to say, the 
rescaling $u(t)$, and therefore the kinetic term of $\sigma$, must remain unchanged. A potential 
term $V(\sigma)$ including $\pi$ or its derivative must therefore be included. Moreover it should 
induce a correction to $\mathfrak{a}^f$ that is higher order than $1/t^2$, so that far from $t=0$ 
(while the matter field is inflating due to the effect of fake de Sitter) the $\sigma_k$ modes that are 
frozen out do not vary with time. The higher order correction in $1/t$ to $\mathfrak{a}^f$ will ensure 
that such an effect only arises during preheating.

We chose to consider a term of the form $e^{2n\pi}\partial_\mu\pi\partial^{\mu}\pi\sigma^2$ with 
$n\geq2$, but a term $e^{2n\pi}\sigma^2$ with $n\geq2$ would have the same effect. Even though 
such terms explicitly break the Galileon symmetry, it is in a very mild way as $t\rightarrow-\infty$ 
since $\sigma$ starts out at zero. Moreover, we expect any coupling of the Galileon to an extra 
degree of freedom to break the Galileon symmetry, and so if we aim to use the Galileon to build 
a cosmological model, we must expect having to break this symmetry in one way or another.

We therefore consider the following  action for the $\sigma$ field interacting with $\pi$:
\be
	\mathcal{S}_I \, = \, \int d^4x\sqrt{-g}\left(-e^{2\pi}\partial_{\mu}\sigma\partial^{\mu}\sigma
	- e^{2n\pi}\partial_{\mu}\pi\partial^{\mu}\pi \sigma^2\right) \, .
\ee
The field rescaling $u(t)$ remains as before. The equation of motion for the rescaled field 
$\tilde\sigma_k(t)$ in Fourier space becomes:
\be
	\ddot{\tilde\sigma}_k + 
	\left(k^2-\frac{2}{t^2}-H_0^2\left(\frac{1}{H_0 t}\right)^{2n}\right)\tilde\sigma_k \, = \, 0 \, .
\ee
With the new coupling, the definition of the fake scale factor $\mathfrak{a}^f$ becomes 
\be
\frac{\ddot{\mathfrak{a}}^f}{\mathfrak{a}^f} \, = \, \frac{\ddot{u}(t)}{u(t)} + V_{,\sigma}
\, = \, \frac{2}{t^2} + \frac{H_0^2}{(H_0t)^{2n}} \, .
\ee
We now see that it will no longer be equal to the field rescaling $u(t)$, which explains the choice 
of different notations for these two variables. 

In what follows, we fix $n=2$, the minimum required 
value. $\mathfrak{a}^{f}$ has an analytical solution which can be approximated in the limits 
$t \rightarrow -\infty$
during the fake de Sitter regime, which we call the IR regime (since it corresponds to the solution 
valid when the IR end of the $\tilde\sigma_k$ spectrum freezes out), and $t \rightarrow 0$ during
the preheating phase, which we call the UV regime (since it corresponds to an approximation valid 
when the UV end of the $\tilde\sigma_k$ spectrum freezes out). The initial conditions are chosen 
such that $\mathfrak{a}^f, \dot{\mathfrak{a}}^f\rightarrow0$ as $t\rightarrow-\infty$:
\bea
	\mathfrak{a}^f(t)	&=& 3H_0 t\left[H_0 t\sinh\left(\frac{1}{(H_0t)}\right)-\cosh\left(\frac{1}{(H_0t)}\right)\right]  \nonumber \\
	&=&-3\sum_{n=0}^{\infty}\frac{1/(H_0t)^{2n+1}}{(2n+1)!(2n+3)} \nonumber \\
	&& \underbrace{\rightarrow}_{t\rightarrow-\infty}~~ \mathfrak{a}^f_{IR}(t)=-\frac{1}{H_0t}\\
	\texttt{and} && \underbrace{\rightarrow}_{~t~\rightarrow~0~} ~~\mathfrak{a}^f_{UV}(t)=-3H_0te^{-1/(H_0t)}\left(\frac{H_0 t+1}{2}\right)
\eea
%%
%I can make the UV solution look simpler by removing the H_0t in the fraction at the end, ie $\mathfrak{a}^f_{UV}(t)=-\frac{3}{2}H_0te^{-1/(H_0t)}$. However it becomes valid much later and is less accurate, i.e around t\sim H_0^{-1}/100. Should I do it, or leave it the way it is?
The overall normalization of $\mathfrak{a}^f$ is chosen such that the IR limit of the spectrum matches 
the corresponding expression in the case of minimal coupling. The approximate solution in the IR
remains  valid up to $t \sim -H_0^{-1}$, at which point the extra term in the differential equation for 
$\mathfrak{a}^f$ starts to dominate. It can then be interpolated with the asymptotic solution in the UV,
which starts being a good approximation at $t\gtrsim -H_0^{-1}/2$. The time of freeze-out of a 
mode $k$ will now be given by:
\bea
	t_{frz} \, = \, -\sqrt{\frac{1}{k^2}+\frac{1}{k^2}\sqrt{1+k^2/H_0^2}}~~&\underbrace{\rightarrow}_{k\rightarrow0}& ~~t_{frz}^{IR}=-\frac{\sqrt{2}}{k} \\
	\texttt{and}~~&\underbrace{\rightarrow}_{k\rightarrow\infty}& ~~t_{frz}^{UV}=-\frac{1}{\sqrt{H_0k}}
\eea
 \begin{figure}[t]
\begin{center}
\includegraphics[scale=0.6]{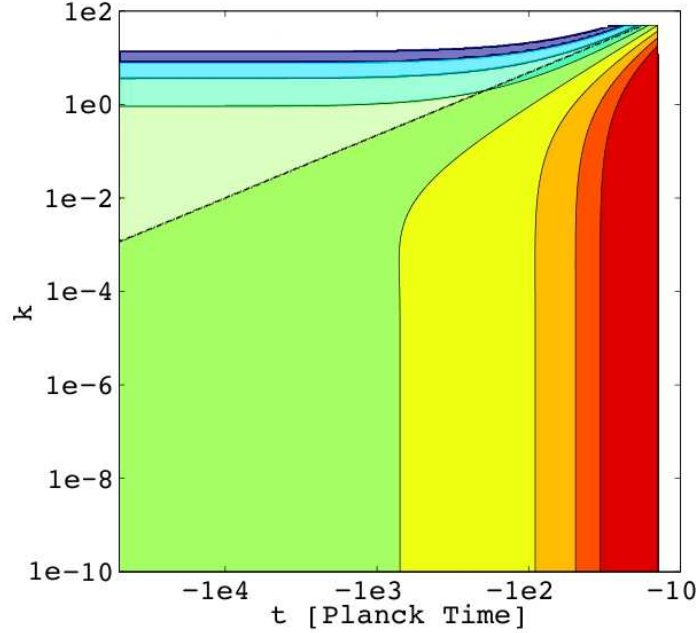}
\end{center}
\caption{Evolution of the power spectrum of the $\sigma_k$ perturbations after freeze-out. The 
level curves show constant amplitudes of perturbations. On the colour map, blue means 
smaller amplitude and red means larger amplitude.  Modes bellow the dotted line are frozen out, 
while modes with paler colour are still oscillating. If we choose for a chosen value of 
$H_0 = 5\times10^{-4}$, the green region corresponds to an amplitude of perturbation of 
$\sim10^{-5}$. The horizontal axis shows the time-evolution of the perturbations for a given k mode, 
while the vertical axis shows the k spectrum at a given time. In the IR end (i.e. for $k<H_0$), the 
spectrum is scale invariant and constant until the onset of preheating after $t = -H_0^{-1}$, at which 
point the amplitude of the fluctuations starts to grow. On the other side, the UV end of the spectrum 
is slightly red and its amplitude increases slightly with time.}
\label{sigmaNonMin-tk}
\end{figure}
The solution for $\sigma_k$ in the regime  $k < k_{frz}$ is therefore:
\bea
	\sigma_k(t) \, = \,  e^{-\pi_{dS}}\frac{\mathfrak{a}^f(t)}{\mathfrak{a}^f(t_{frz})}\frac{1}{\sqrt{2k}}~~&\underbrace{\rightarrow}_{k\rightarrow0}& ~~\sigma_k^{IR}\simeq\begin{cases} \frac{H_0}{k^{3/2}} ~~ ,~|t|\gtrsim H_0^{-1} \\ \\
		\frac{3H_0^3t^2}{k^{3/2}}e^{-1/(H_0t)}(\frac{H_0t+1}{2})  ~~ , ~0<|t|\lesssim H_0^{-1}/10\end{cases}  \\\nonumber\\
		\label{sigUVnonmin}
		\texttt{and}~~&\underbrace{\rightarrow}_{k\rightarrow\infty}& ~~\sigma_k^{UV}\simeq\frac{H_0^2t^2k^{1/2}}{\sqrt{2}}e^{-\frac{1+t\sqrt{H_0k}}{H_0t}}\frac{H_0t+1}{\sqrt{H_0k}-H_0}
\eea
Figure \ref{sigmaNonMin-tk} shows the $\sigma_k$ power spectrum as a function of time. We 
see that, upon the addition of the extra term in the action, the IR end of the 
$\sigma_k$ spectrum stays scale invariant and remains constant after freeze-out (which is the 
condition for the solution to be an attractor) until the time of preheating, which happens at 
$|t| \lesssim H_0^{-1}$. Modes freezing out very close to reheating will not be characterized 
by a scale-invariant spectrum, but rather by a red spectrum, as can be seen from 
(\ref{sigUVnonmin}). However these modes are outside of the observable range since they 
re-enter the oscillatory regime just after preheating and will not have any observable 
consequences to leading order. However it might be interesting to see whether these small scales 
can couple together to have effects on larger scales that are observable today, in the form of the
introduction of non-Gaussianity,  for example.

Repeating the analysis presented above to get the energy density in $\rho_\sigma$, it is easy to 
see that one obtains an extra term in the stress-energy for $\sigma$ that will add an extra 
$+\frac{1}{t^2}\frac{1}{(tH_0)^{4}}\sigma^2$ to the energy density in each $k$ mode. Integrating 
as above, we obtain (being careful to separate properly the regions of $k$ space and the 
temporal regions of validity of our approximate solutions)
 \be
 	\bar{\rho}_\sigma \, \simeq \, \int_{k_i}^{k} \frac{d^3 k}{(2\pi)^3} \left( \frac{1}{k t^2}+\frac{1}{t^6H_0^2k^3} \right) = \frac{1}{2\pi^2t^2}\left( \frac{1}{t^2}-\frac{1}{t_i^2} \right)+\frac{1}{2\pi^2 t^6 H_0^2}ln\left[ \frac{t_i}{t} \right] \qquad \texttt{for }|t|\gtrsim H_0^{-1}\texttt{ and }k<H_0 \, ,
 \ee
during the IR regime, or fake de Sitter epoch. At $t \sim-H_0^{-1}$, we enter the UV regime: the 
behaviour of the modes that froze out during the IR regime changes and the modes that freeze  
out have a different time evolution. It is at this time that the defrosting/ preheating starts, and 
$\bar{\rho}_\sigma$ will now be given by:
 \bea
 	\bar{\rho}_\sigma \, &\simeq& \, \int_{k_i}^{H_0} \frac{d^3 k}{(2\pi)^3}\rho^{IR}(\mathbf{k},t)+\int_{H_0}^{k} \frac{d^3 k}{(2\pi)^3}\rho^{UV}(\mathbf{k},t)\nonumber \\
	&=& \, \frac{H_0^2}{2\pi^2}e^{-2/H_0t}\left[\left( \frac{H_0^2}{2}-\frac{1}{t_i^2} \right)\mathcal{O}(H_0^4t^4)+ln\left(\frac{H_0t_i}{\sqrt{2}}\right)\mathcal{O}\left(\frac{1}{t^2}\right)\right]+\frac{1}{2\pi^2}e^{-1/H_0t}\left[ \mathcal{O}\left(\frac{1}{H^{12}t^{16}}\right) \right]\\
	&& \qquad\qquad \qquad \qquad \qquad \qquad \qquad \qquad \qquad \qquad\qquad \qquad \qquad \qquad \qquad \texttt{for }|t|< H_0^{-1}\texttt{ and }k>H_0 \, ,\nonumber
%	the first integral is: &&\qquad+\frac{H_0^2}{2\pi^2}\frac{9}{4}e^{-2/H_0t}\left[ \left[ \frac{1}{t^2}-\frac{H_0^2}{2}\right](H_0t)^4(H_0t+1)^2+ln\left(\frac{\sqrt{2}}{H_0t}\right)\mathcal{O}\left(\frac{1}{t^2}\right)\right] \nonumber \\
 \eea
where we have only included the lowest order terms in $t$ in the result for clarity. With the new 
coupling, $\bar{\rho}_\sigma$ grows sufficiently fast when $|t| < H^{-1}_0$ to overcome even a more 
realistic estimate of the growth rate of $\rho_\pi$ which diverges as $\sim1/t^6$ as 
$t\rightarrow0$. Hence the growth of the energy density in the matter field will be fast enough to  allow for initiation of preheating.
 
It only remains to check whether the introduction of the new coupling term back-reacts on the 
$\pi$ background in a way as to slow it down, or if it spoils the relationship we had previously 
obtained. However, it is straightforward to check that the only modification is that equation (\ref{homobackreact2}) acquires the extra source term on the r.h.s: 
\be
\frac{2t\delta\sigma\delta\dot\sigma-3\delta\sigma^2}{f^2H_0^2t^4} \, .
\ee
Since we know from above (and as can easily be seen from Figure \ref{sigmaNonMin-tk}) that 
$\delta\sigma$ and $\delta\dot\sigma$ are always positive, and since $t < 0$, the contribution 
of this extra term is always {\it negative}. Hence the source term back-reacting on the 
homogeneous mode of the $\pi$ background becomes more negative compared to the minimal 
coupling case, which means that the Galileon will be even more efficiently driven toward the 
$\pi=0$ solution and preheating will be completed on a more rapid time scale.

Qualitatively, as the Galileon rolls toward $t=0$, it excites fluctuations in the matter field. From the 
point of view of the matter field, it gets excited in exactly the same manner as if it was immersed in 
de Sitter space. However, instead of getting the energy for particle production from the metric, it is 
the Galileon that transfers it through its coupling to matter. Hence as fluctuations in the matter field get 
amplified, the Galileon is slowed down. By adding an extra coupling term, the energy transfer 
toward the matter field was made much more efficient starting at $|t|<H_0^{-1}$, hence the 
$\sigma$ growth was accelerated and $\pi$ was accordingly decelerated, making preheating possible.

If, instead, we had chosen to add a coupling term of the form $e^{2n\pi}\sigma^2$ with $n=2$, the 
effect on $\sigma$ would have been qualitatively the same and the only difference on the $\pi$ 
back-reaction is that the extra source term would have been 
\be
-\frac{4\delta\sigma^2}{f^2H_0^2t^2} \, .
\ee
Hence the Galileon would still have been slowed down, but in a less efficient way than with the coupling considered above. If we had chosen $n>2$, the energy transfer to $\sigma$ and the resulting deceleration back-reaction on $\pi$ would have been even faster.

\section{Reintroducing the Coupling to Gravity}
\label{Hneq0}

In the above section, it was discussed how an additional degree of freedom, which we took to 
be a matter scalar field, behaves when coupled both minimally and non-minimally to the Galileon 
in its de Sitter solution, when fields are decoupled from gravity. It was realized that in the case of 
minimal kinematic coupling between $\pi$ and $\sigma$, the energy density in $\sigma$ does not 
grow fast enough to allow for the onset of preheating. However, it was shown how adding extra 
coupling between the Galileon and the matter field can allow us to overcome this difficulty.

In what follows, we study what happens if we re-introduce the coupling of the fields to gravity at the 
level of the background. As discussed earlier, Minkowski space and $\pi_{dS}$ are not a solution 
to Einstein's equations anymore, but they are only asymptotically in the past. We therefore use
(\ref{piwGrav-tinfinity}) and (\ref{piwGrav-tclosesing}) as a background and revisit the case of minimal
coupling with $\sigma$. Surprisingly, it turns out that including this new background will have a 
very similar effect as the extra coupling we added in the previous subsection. Hence, contrary to what
one might expect, including the effects of $H$ in the background will {\it accelerate} the $\sigma$ field. 

We again consider the action (\ref{Actionminimalcoupling}), but we now assume an FRW metric to 
derive the  equation of motion for $\sigma$. Asymptotically as $t\rightarrow-\infty$, we still have 
$t=\tau$. However, the conformal time is no longer equal to the physical time at any finite time. 
We therefore need to cast the equation of motion for $\sigma$ in conformal time to find the proper
canonically-normalized variable. We find that
\be
u \, = \, e^{\pi}a
\ee
is required in order to transform the kinetic term of $\sigma$ into a flat-space canonical form. Upon 
the field rescaling 
\be
\sigma \, \rightarrow \, u^{-1}\tilde\sigma \, , 
\ee
we obtain the Fourier space equation of motion
\be
\label{SigGravEoM}
	\tilde\sigma_k'' + \left(k^2-\pi''-(\pi'+\mathcal{H})^2-\mathcal{H}'\right)\tilde\sigma_k \, = \, 0 \, .
\ee
Here, primes denote derivatives with respect the conformal time and $\mathcal{H}$ stands for 
the conformal Hubble factor, $a'/a$. It is easy to verify that the fake scale factor $\mathfrak{a}^f$ 
satisfies $(\mathfrak{a}^f)''/\mathfrak{a}^f = u''/u$, which ensures that the amplitudes of the 
$k$-modes remain constant after their freeze-out. More precisely, for $k < k_{frz}$,
\be
\label{sigmaGrav}
	\sigma_k \, = \, u^{-1}(t)\frac{(\mathfrak{a}^f(t_{frz}))^{-1}}{(\mathfrak{a}^f(t))^{-1}}\tilde\sigma(t_i)
	\, = \, e^{-\pi(t_{frz})}a^{-1}(t_{frz})\frac{1}{\sqrt{2k}}.
\ee
Since we know the background solution in terms of physical time, it is easier to work in terms of 
$t$ instead of $\tau$. 

In order to find the explicit form of the spectrum of perturbations, a solution for $t_{frz}$ is 
required for both asymptotic background regimes: (\ref{piwGrav-tinfinity}) as $t\rightarrow-\infty$ 
and (\ref{piwGrav-tclosesing}) as $t\rightarrow t_0$. Starting by analysing the first regime, we write 
the explicit form of the time-dependent mass term in (\ref{SigGravEoM}) in terms of the physical time 
as follows:
\bea
	\frac{(\mathfrak{a}^f)''}{\mathfrak{a}^f} \, &=& 
	\, a^2\left[ \ddot\pi+\dot\pi^2 +\dot\pi H+2H^2 +\dot{H}\right]\nonumber\\
	&\underbrace{\rightarrow}_{t\rightarrow-\infty}& \, 
	e^{\frac{1}{3}\frac{f^2}{m_{pl}^2}\frac{1}{H_0^2t^2}}\left[\frac{2}{t^2}-\frac{2}{3}\frac{f^2}{m_{pl}^2}\frac{1}{H_0^2t^4}+2H^2+\frac{8}{9}\frac{f^4}{m_{pl}^4}\frac{1}{H_0^4t^6}\right] \, .
	\label{tfrzGravIR}
\eea
Evaluating (\ref{tfrzGravIR}) at $t_{frz}$ and setting the whole equation equal to $k^2$ yields the 
relationship between the time of freezeout and the wavenumber of a mode which we need.

The region where the background (\ref{piwGrav-tinfinity}) is valid extends up to $t\sim-H_0^{-1}f/m{pl}$, 
and hence matches what we earlier called the IR region. Since $a$ starts out equal to 1 as 
$t\rightarrow-\infty$, equation (\ref{tfrzGravIR}) is dominated by the $2/t^2$ over the whole region, 
so that $t_{frz}\simeq-\sqrt{2}/k$ for all times when $a\approx1$, just as in the previous section. This ensures that this region the spectrum is quasi-scale invariant:
\be
	\sigma^{IR}_k \, = \, \frac{H_0}{k^{3/2}}e^{\frac{1}{6}\frac{f^2}{m^2_{pl}}\frac{k^2}{H_0^2}}\qquad k<k_{frz}\lesssim H_0 m_{pl}/f,~ -H_0^{-1}f/m_{pl}\gtrsim t>t_{frz} \, .
\ee
We now look at the growth of energy density in the matter field. From (\ref{stress-energy}), the energy
density in each frozen $k$ mode is given by
\be
	\rho^{IR}(\mathbf{k}, t) \, \simeq \, \frac{e^{-\frac{4}{3}\frac{f^2}{m_{pl}^2}\frac{1}{H_0^2t^2}}}{(tH_0)^2}\left[\frac{H_0^2}{k}e^{\frac{1}{3}\frac{f^2}{m_{pl}^2}\frac{k^2}{H_0^2}}\right]
	 \qquad k<k_{frz}\lesssim H_0 m_{pl}/f,~ -H_0^{-1}f/m_{pl}\gtrsim t>t_{frz} \, .
\ee
The energy density $\bar{\rho}_\sigma$ during the IR regime will therefore be given by:
\be
	\bar{\rho}_\sigma^{IR} \, = \, \int_{k_i}^{k} \frac{d^3 k}{(2\pi)^3} \rho^{IR}(\mathbf{k}, t) \simeq \frac{3}{4\pi^2}\frac{H_0^2}{t^2}\frac{m_{pl}^2}{f^2}\left[\exp\left[-\frac{2}{3}\frac{f^2}{m_{pl}^2}\frac{1}{H_0^2t^2}\right]-\exp\left[\frac{2}{3}\frac{f^2}{m_{pl}^2}\frac{1}{H_0^2}\left(\frac{1}{t_i^2}-\frac{2}{t^2}\right)\right]\right] 
	\qquad t\lesssim-H_0^{-1}f/m_{pl} \, .
\ee
As expected, Taylor expanding the exponential, we recover the gravity-decoupled result
(\ref{rhoSigdeGravMin}) in the limit when $t\rightarrow-\infty$. 

As $t\approx -H^{-1}$, the background solution shifts smoothly from (\ref{piwGrav-tinfinity}) to 
(\ref{piwGrav-tclosesing}), and $\sigma$ enters a new regime of evolution that we call the UV 
regime. Now, the time-dependent mass is given by
\bea
\label{UVgravTimeDepMass}
	\frac{(\mathfrak{a}^f)''}{\mathfrak{a}^f}&\underbrace{\rightarrow}_{t\rightarrow t_0}& e^{\frac{16}{3}\frac{f^2}{m_{pl}^2}\frac{1}{H_0^2(t_0-t)^2}}\left[\frac{6}{(t_0-t)^2}+\frac{80}{3}\frac{f^2}{m_{pl}^2}\frac{1}{H_0^2(t_0-t)^4}+\frac{512}{9}\frac{f^4}{m_{pl}^4}\frac{1}{H_0^4(t_0-t)^6}\right] \nonumber \\
	&&\approx e^{\frac{16}{3}\frac{f^2}{m_{pl}^2}\frac{1}{H_0^2(t_0-t)^2}}\left[\frac{512}{9}\frac{f^4}{m_{pl}^4}\frac{1}{H_0^4(t_0-t)^6}\right] \, .
\eea
In the last step, we have used the fact that the term $\frac{6}{(t_0-t)^2}$ has already stopped 
being dominant at the time when the background enters the UV regime, and we have also 
neglected the $\sim\frac{1}{H_0^4(t_0-t)^4}$ term since we are interested in the behaviour 
of $\sigma$ as the background gets close to the singularity at $t_0$.

Working in this approximation, the freeze-out time in the UV regime is given by a Lambert W 
function $W_0$ of the comoving wavenumber, which cannot be expressed in terms of elementary 
functions and whose divergence at infinity is slower than the divergence of a logarithmic function:
\be
	(t_0-t_{frz}) \, \simeq \, \frac{4}{3H_0}\frac{1}{\sqrt{W_0\left(\frac{2}{3^{4/3}}\left(\frac{k}{H}\right)^{2/3}\right)}},  \qquad t\rightarrow t_0 \, .
\ee
Inserting this result into (\ref{sigmaGrav}), we obtain the approximate solution for $\sigma_k$ in the 
UV regime:
\bea
	k_{frz}>k\gtrsim H_0 m_{pl}/f,~ t>t_{frz}\gtrsim-H_0^{-1}f/m_{pl}, \qquad\sigma^{UV}_k&\simeq&\frac{m_{pl}}{f}\frac{\sqrt{2}e^{-\frac{3}{2}\frac{f^2}{m_{pl}^2}W_0\left(\frac{2}{3^{4/3}}\left(\frac{k}{H}\right)^{2/3}\right)}}{3\sqrt{3k}W_0\left(\frac{2}{3^{4/3}}\left(\frac{k}{H}\right)^{2/3}\right)} \nonumber \\
	&=&\frac{m_{pl}}{f}\frac{H_0\sqrt{3}}{2 k^{3/2}}\left[W_0\left(\frac{2}{3^{4/3}}\left(\frac{k}{H}\right)^{\frac{2}{3}}\right)\right]^{\left(\frac{3}{2}\frac{f^2}{m_{pl}^2}-1\right)} 
\eea
Hence we see that as the matter field enters the UV regime, the power spectrum of $\sigma_k$
perturbations starts to deviate from scale independence and is multiplied by an additional divergent part 
that grows as the square of a Lambert W function as the background approaches the singularity. 
This means that in the UV regime the spectrum gets more and more tilted toward the blue as 
$t\rightarrow t_0$. Therefore, not only does the re-introduction of the coupling to gravity in the 
background push the singularity forward in time from $t = 0$ to 
$t = t_0\sim\frac{3}{2}\frac{f}{m_pl}H_0^{-1}$, but, quite surprisingly, as $\sigma$ approaches 
the singularity, its perturbations grow {\it faster} than when we fix $H=0$.

 \begin{figure}[t]
\begin{center}
\includegraphics[scale=0.4]{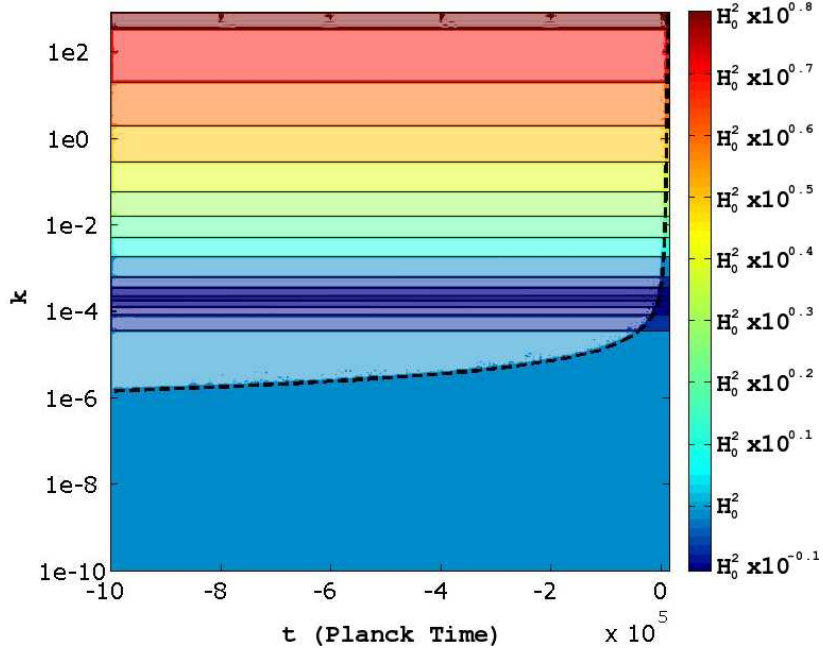}
\end{center}
\caption{Power spectrum of $\sigma_k$ fluctuations after freezeout when the full background 
with gravitational coupling is considered, obtained by solving numerically the differential equation f
or the background $\pi$ and $H$ and the time-dependent mass in (\ref{UVgravTimeDepMass}). 
The colour map is logarithmic in the amplitude of the power spectrum, and increases from blue to red.
The horizontal axis shows the time evolution of the perturbations for a given k mode, while the vertical
axis shows the logarithm of the k spectrum at a given time. Modes below the dotted line are frozen 
out with the indicated amplitude, while modes with paler colour above the dotted line are still oscillating.
The first thing to note is that the spectrum is now constant for all modes that are frozen out. At the IR 
end (i.e. for $k\lesssim H_0$), the spectrum is almost scale-invariant and slightly red-tilted until the end 
of that regime after $t=-H_0^{-1}$. On the other side, the UV end of the spectrum is very tilted towards 
the blue, and the onset of this UV regime marks the onset of preheating. In between the two regimes, 
the graph shows an interesting feature in the power spectrum: a small trough between $k=10^{-4}$ 
and $k=10^{-3}$. This feature allows for a red-tilted power spectrum during the fake de Sitter phase, 
or IR regime, where the spectrum is almost scale invariant, at the same time as an efficient preheating
with a blue spectrum in the UV end, so that the power spectrum becomes highly dominated by  
UV modes as defrosting proceeds.}
\label{sigmaGRAV}
\end{figure}

If we find the full numerical solution for the background Galileon and Hubble rate, as well as 
for the time-dependent mass (\ref{UVgravTimeDepMass}) and insert them into (\ref{sigmaGrav}), 
we obtain solutions for $\sigma_k$ for every $k$ after their freeze-out and can compute the full 
power spectrum of $\sigma_k$ perturbations. The result is shown in 
Figure \ref{sigmaGRAV} for $f=m_{pl}$ and $H_0=10^{-4}$. One 
interesting feature of the power spectrum is the small trough between $k=10^{-4}$ and $k=10^{-3}$.
Because of this feature, the power spectrum in the scale invariant region (on scales larger than
$k^{-1}=10^{5}$) is slightly {\it red tilted} up to scales where it reaches the bottom of the trough. 
For larger values of $k$ it then becomes very heavily blue tilted.

Turning now to the evaluation of the growth of the energy density during the UV regime, we write 
the energy density in every frozen $k$-mode as:
\bea
	\rho^{IR}(\mathbf{k}, t) &\simeq& \frac{2^6}{3}\frac{f^2}{m_{pl}^2}\frac{e^{-\frac{16}{3}\frac{f^2}{m_{pl}^2}\frac{1}{H_0^2(t-t_0)^2}}}{H_0^4(t_0-t)^4}\left[\frac{H_0^2}{k}e^{\frac{1}{3}\frac{f^2}{m_{pl}^2}\frac{k^2}{H_0^2}}\right]
	 \qquad k<k_{frz}\lesssim H_0 \frac{m_{pl}}{f},~ t_{frz}<-H_0^{-1}\frac{f}{m_{pl}}\lesssim t \, ; \\
\rho^{UV}_\sigma(\mathbf{k}, t) &\simeq& \frac{2^4e^{-\frac{16}{3}\frac{f^2}{m_{pl}^2}\frac{1}{H_0^2(t-t_0)^2}}}{H_0^2(t_0-t)^4}\frac{\left[W_0\left(\frac{2}{3^{4/3}}\left(\frac{k}{H}\right)^{\frac{2}{3}}\right)\right]^{\left(3\frac{f^2}{m_{pl}^2}-2\right)} }{k} \qquad k>k_{frz}\gtrsim H_0 \frac{m_{pl}}{f},~ -H_0^{-1}\frac{f}{m_{pl}}\lesssim t_{frz}<t\, .
\eea
The total averaged energy density $\bar{\rho}_\sigma$ during the UV regime is now:
\bea
	\bar{\rho}_\sigma^{UV} &=& \int_{k_i}^{\left.k_{frz}\right|_{t=-H_0^{-1}}} \frac{d^3 k}{(2\pi)^3} \rho^{IR}(\mathbf{k}, t) + \int_{\left.k_{frz}\right|_{t=-H_0^{-1}}}^{k} \frac{d^3 k}{(2\pi)^3} \rho^{UV}(\mathbf{k}, t) \\
	&\simeq& \frac{2^4}{\pi^2}\frac{e^{-\frac{16}{3}\frac{f^2}{m_{pl}^2}\frac{1}{H_0^2(t-t_0)^2}}}{(t-t_0)^4}\left[\exp\left[\frac{2}{3}\frac{f^2}{m_{pl}^2}\right]-\exp\left[\frac{2}{3}\frac{f^2}{m_{pl}^2}\frac{1}{H_0^2t_i^2}\right]\right] +\frac{3^5}{2\pi^2}\frac{e^{-\frac{16}{3}\frac{f^2}{m_{pl}^2}\frac{1}{H_0^2(t-t_0)^2}}}{(t-t_0)^4}\int dw (1+w)w^{3\frac{f^2}{m_{pl}^2}}e^{3w} 
	\, , \nonumber 
\eea
where we have used the change of variables $we^{w}=x$ and $W_0(x)=w$ with 
$x=\frac{2}{3^{4/3}}\left(\frac{k}{H_0}\right)^{2/3}$. The limits of integration are still from 
$w$ evaluated at the scale freezing out at the time when the background just enters the UV regime, 
up to $w$ evaluated to the scale freezing out at the time when we want to know the energy density. Considering that a realistic value of $f$ implies $f\sim m_{pl}$ \cite{Nicolis:2008in}, it is realistic to fix $f=m_{pl}$. Making use
of that assumption, the remaining integral can be evaluated, yielding a leading diverging behaviour 
close to the singularity at $t_0$:
\be
	\bar{\rho}_\sigma^{UV}  \, \sim \, \frac{2^{11}}{3^2\pi^2}\frac{W_0\left(\frac{2^4}{3^2}\frac{e^{\frac{2^4}{3^2}\frac{1}{H_0^2(t_0-t)^2}}}{H_0^2(t_0-t)^2}\right)}{H_0^6(t_0-t)^{10}}~~ , ~t\rightarrow t_0 \, .
\ee
We have used (\ref{tfrzGravIR}) to relate $k$-modes to their corresponding time of freeze-out. 
We now see that when the coupling to gravity is re-introduced, the leading contribution to the 
average energy density in $\sigma$, $\bar{\rho}_\sigma$, in the UV regime grows faster 
than $\sim1/(t_0-t)^{10}$, while, 
as discussed before, the energy density in the Galileon, $\rho_\pi$, grows as $1/(t_0-t)^6$. 
This means that when a full treatment of the background including couplings to gravity is 
considered, then as the Galileon rolls close to the singularity at $t_0$, the energy density in the 
matter field with minimal kinetic coupling will eventually come to dominate the Galileon, allowing 
for the onset of preheating. 

The only thing left now for us to check in order to ensure that the Galilean Genesis model does 
not suffer from a graceful exit problem is the back-reaction of $\sigma$ on $\pi$. That is, we need 
to verify that, once coupling to gravity is re-introduced, the growth of $\sigma$ will still slow down 
the Galileon and not accelerate it, and make it evolve from the de Sitter configuration toward its 
$\pi = {\rm constant}$  solution. This, however, is very straightforward to check, since although 
the EoM for $\pi$ (still working in physical time) becomes significantly more complex 
(see \cite{CNT}), the source 
term coming from $\sigma$ is still the same as before and does not involve any additional terms 
in $H$. That is, the source term from $\sigma$ is still: 
\be
-\frac{1}{f^2}\left[\dot{\sigma}^2-\frac{(\nabla\sigma)^2}{a^2} \right] \, . 
\ee

When computing the back-reaction EoM for $\pi$ in Fourier space, we again assume a fixed 
background metric. Indeed, it is reasonable to assume that the time scale for the instability in 
$\sigma$ to develop and overcome the evolution of the background is much shorter than 
the characteristic time scale for any metric perturbations to become important. It is therefore 
possible to write the contribution of $\sigma$ to the background Galileon back-reaction equation 
$\delta^{(2)}\pi_{k=0}$ as a source term on the r.h.s. of the form:
\be
	\Delta~EoM^{r.h.s.}_{\delta^{(2)}\pi_{k=0}} \, = \, -\frac{1}{f^2}e^{-2\pi_{0}}\bar{\rho}_\sigma \, .
\ee
Since $\sigma$ is a regular matter field  $\rho_\sigma$ is obviously always a positive quantity. 
Therefore, the additional source term that $\sigma$ contributes to the back-reaction to the 
Galileon background always has the effect of slowing down $\pi$. A more explicit approximate expression of this source term can be obtained in each of the two regimes considered above 
for $\pi_0$ and $\rho_\sigma$. In the IR regime, we obtain:
\bea
	\Delta^{IR}~EoM^{r.h.s.}_{\delta^{(2)}\pi_{k=0}}\approx -\frac{3H_0^4}{4\pi^2}\frac{m_{pl}^2}{f^4}\left[e^{\frac{1}{3}\frac{f^2}{m_{pl}^2}\frac{1}{H_0^2t^2}}-e^{\frac{1}{3}\frac{f^2}{m_{pl}}^2\frac{1}{H_0^2}(\frac{2}{t_i^2}-\frac{1}{t^2})}\right] ~~ ,~t\lesssim -H_0^{-1}f/m_{pl} \, ,
\eea
while in the UV regime, setting $m_{pl}=f$, we obtain that the leading contribution to the source 
term is:
\bea
		\Delta^{UV}~EoM^{r.h.s.}_{\delta^{(2)}\pi_{k=0}}\approx-\frac{2^5}{3\pi^2}\frac{1}{f^2}\frac{W_0\left(\frac{2^4}{3^2}\frac{e^{\frac{2^4}{3^2}\frac{1}{H_0^2(t_0-t)^2}}}{H_0^2(t_0-t)^2}\right)}{H_0^2(t_0-t)^6}  ~~ , ~t\rightarrow t_0 \, .
\eea
We can therefore conclude that, as fluctuations in the matter field get amplified because of their immersion in fake de Sitter space, and as the total energy density in $\sigma$ grows accordingly, 
the Galileon from which $\sigma$ gets its energy slows down. This therefore ensures that the Galilean 
Genesis scenario does not suffer from a graceful exit problem, and that the system will proceed 
to a FRW phase dominated by $\sigma$, in which the NEC is re-established with $\dot{H}<0$. 

\section{Conclusions and Discussion}

In this paper we have studied the transition between an Emergent Galileon background
phase and the radiation phase of an expanding universe. We have shown that, at least
when including the effects of non-vanishing expansion on the background fields, the
same coupling of the Galileon condensate to regular scalar field matter introduced in
\cite{CNT} is sufficient to ensure a rapid energy transfer to the matter field which then via
its back-reaction on the Galileon background leads to a slowing down of the Galileon
condensate. 

There are similarities and differences between the defrosting transition of the Galileon
background studied here and preheating in inflationary cosmology. In both cases,
it is the coherent dynamics of the background matter field which drives the production
of regular matter. Here it is the dynamics of the Galileon background, in inflationary
cosmology it is the coherent oscillations of the inflaton condensate at the end of the
period of inflation. However, here it is the same squeezing of fluctuations which leads 
to scale-invariant matter fluctuations which leads to defrosting/preheating, whereas in 
inflationary cosmology the generation of scale-invariant fluctuations and the reheating 
instability are separate processes.

In inflationary cosmology, it is mostly long wavelength modes which are excited during
preheating. On the other hand, we have shown here that efficient Galileon defrosting
is based on a sharp blue tilt of the spectrum in the UV. In light of these similarities
and differences it would be of great interest to study reheating in Galileon-based
inflation models \cite{Galileoninflation}.

We wish to end with a comment on the generation of curvature fluctuations in the
Emergent Galileon scenario: since the background matter has vanishing background
value, a scale-invariant spectrum of the matter fields does not lead to scale-invariant
spectrum of curvature fluctuations (see e.g. \cite{Durrer:1994}) since the curvature
fluctuations are quadratic in the matter perturbations. In addition, if the matter fluctuations
have Gaussian statistics, the curvature perturbations with not be Gaussian. This is
an issue which merits further study.

\begin{acknowledgments}

The research of R.B. is supported in part by an NSERC Discovery Grant at 
McGill University and by funds from the Canada Research Chairs program; the research of ACD is supported in part by STFC. The research of L.P.L is supported in part by an FQRNT B1 scholarship. R.B. wishes to acknowledge the hospitality of Professor Xinmin Zhang and the Theory Division of the Institute of High Energy Physics in Beijing for hospitality
while these results were written up. ACD wishes to thank the Department of
Physics, McGill University for hospitality whilst this research was initiated.

\end{acknowledgments}


\begin{thebibliography}{99}

%1\cite{Riess:1998cb}
\bibitem{Riess:1998cb}
  A.~G.~Riess {\it et al.}  [Supernova Search Team Collaboration],
  %``Observational evidence from supernovae for an accelerating universe and a
  %cosmological constant,''
  Astron.\ J.\  {\bf 116}, 1009 (1998)
  [arXiv:astro-ph/9805201].
  %%CITATION = ANJOA,116,1009;%%
  
%2\cite{Perlmutter:1998np}
\bibitem{Perlmutter:1998np}
  S.~Perlmutter {\it et al.}  [Supernova Cosmology Project Collaboration],
  %``Measurements of Omega and Lambda from 42 high redshift supernovae,''
  Astrophys.\ J.\  {\bf 517}, 565 (1999)
  [arXiv:astro-ph/9812133].
  %%CITATION = ASJOA,517,565;%%

%3\cite{Weinberg:1987dv}
\bibitem{Weinberg:1987dv}
  S.~Weinberg,
  %``Anthropic Bound on the Cosmological Constant,''
  Phys.\ Rev.\ Lett.\  {\bf 59}, 2607 (1987).
  %%CITATION = PRLTA,59,2607;%%

%4\cite{Khoury:2010xi}
\bibitem{Khoury:2010xi}
  J.~Khoury,
  %``Theories of Dark Energy with Screening Mechanisms,''
  arXiv:1011.5909 [astro-ph.CO].
  %%CITATION = ARXIV:1011.5909;%%

%5
\bibitem[Upadhye et al.(2006)]{2006PhRvD..74j4024U}  
A.~Upadhye, S.~S.~Gubser and J.~Khoury, \prd {\bf 74}, 104024 (2006).

%6\cite{Brax:2004px}
\bibitem{Brax:2004px}
  P.~Brax, C.~van de Bruck, A.~C.~Davis, J.~Khoury and A.~Weltman,
  %``Chameleon dark energy,''
  AIP Conf.\ Proc.\  {\bf 736}, 105 (2005)
  [arXiv:astro-ph/0410103].
  %%CITATION = APCPC,736,105;%%

%7
\bibitem[Brax et al.(2004)]{2004PhRvD..70l3518B} 
P.~Brax, C.~van de Bruck, A.-C.~Davis, J.~Khoury and A.~Weltman, \prd {\bf 70}, 123518 (2004).

%8
\bibitem[Gubser
\& Khoury(2004)]{2004PhRvD..70j4001G} 
S.~S.~Gubser and J.~Khoury, \prd {\bf 70}, 104001 (2004).

%9
\bibitem[Khoury
\& Weltman(2004)]{2004PhRvL..93q1104K} 
J.~Khoury and A.~Weltman, Phys.\ Rev.\ Lett.\ {\bf 93}, 171104 (2004).

%10
\bibitem[Khoury
\& Weltman(2004)]{2004PhRvD..69d4026K} 
J,~Khoury and A.~Weltman, \prd {\bf 69}, 044026 (2004).

%11
\bibitem[Hinterbichler
\& Khoury(2010)]{2010PhRvL.104w1301H} 
K.~Hinterbichler and J.~Khoury, Phys.\ Rev.\ Lett.\ {\bf 104}, 231301 (2010).

%12
\bibitem[Olive
\& Pospelov(2008)]{2008PhRvD..77d3524O} 
K.~A.~Olive and M.~Pospelov, \prd {\bf 77}, 043524 (2008).

%13
\bibitem[Pietroni(2005)]{2005PhRvD..72d3535P} 
M.~Pietroni, \prd {\bf 72}, 043535 (2005).

%14\cite{Vainshtein:1972sx}
\bibitem{Vainshtein:1972sx}
  A.~I.~Vainshtein,
  %``To the problem of nonvanishing gravitation mass,''
  Phys.\ Lett.\  B {\bf 39}, 393 (1972).
  %%CITATION = PHLTA,B39,393;%%

%15
\bibitem[Arkani-Hamed et al.(2003)]{2003AnPhy.305...96A} 
N.~Arkani-Hamed, H.~Georgi and M.~D.~Schwartz, Annals of Physics {\bf 305}, 96 (2003).

%16
\bibitem[Deffayet et al.(2002)]{2002PhRvD..65d4026D} 
C.~Deffayet, G.~Dvali, G.~Gabadadze and A.~Vainshtein, \prd {\bf 65}, 044026 (2002).

%17
\bibitem[de Rham \& Gabadadze(2010)]{2010PhRvD..82d4020D} 
C.~de~Rham and G.~Gabadadze, \prd {\bf 82}, 044020 (2010).

%18
\bibitem[de Rham(2010)]{2010PhLB..688..137D}  
C.~de~Rham, Phys.\ Lett.\  B {\bf 688}, 137 (2010).

%19
\bibitem[Gabadadze(2009)]{2009PhLB..681...89G} 
G.~Gabadadze, Phys.\ Lett.\  B {\bf 681}, 89 (2009).

%20
\bibitem[Dvali et al.(2007)]{2007PhRvD..76h4006D} 
G.~Dvali, S.~Hofmann and J.~Khoury, \prd {\bf 76}, 084006 (2007).

%21
\bibitem[Dvali et al.(2003)]{2003PhRvD..67d4020D} 
G.~Dvali, G.~Gabadadze and M.~Shifman, \prd {\bf 67}, 044020 (2003).

%22
\bibitem[Arkani-Hamed et al.(2002)]{2002hep.th....9227A} 
N.~Arkani-Hamed, S.~Dimopoulos, G.~Dvali and G.~Gabadadze [arXiv:hep-th/0209227].

\bibitem{Patil}
S.~P.~Patil,
  %``Degravitation, Inflation and the Cosmological Constant as an Afterglow,''
  JCAP {\bf 0901}, 017 (2009)
  [arXiv:0801.2151 [hep-th]].
  %%CITATION = JCAPA,0901,017;%%
  
%23\cite{Dvali:2000hr}
\bibitem{Dvali:2000hr}
  G.~R.~Dvali, G.~Gabadadze and M.~Porrati,
  %``4-D gravity on a brane in 5-D Minkowski space,''
  Phys.\ Lett.\  B {\bf 485}, 208 (2000)
  [arXiv:hep-th/0005016].
  %%CITATION = PHLTA,B485,208;%%

%24\cite{Deffayet:2001pu}
\bibitem{Deffayet:2001pu}
  C.~Deffayet, G.~R.~Dvali and G.~Gabadadze,
  %``Accelerated universe from gravity leaking to extra dimensions,''
  Phys.\ Rev.\  D {\bf 65}, 044023 (2002)
  [arXiv:astro-ph/0105068].
  %%CITATION = PHRVA,D65,044023;%%

%25
\bibitem[de Rham et al.(2010)]{2010PhRvD..81l4027D} 
C.~de~Rham, J.~Khoury and A.~J.~Tolley , \prd {\bf 81}, 124027 (2010).

%26
\bibitem[Agarwal et al.(2010)]{2010PhRvD..81h4020A} 
N.~Agarwal, R.~Bean,
J.~Khoury and M.~Trodden, \prd {\bf 81}, 084020 (2010).

%27
\bibitem[de Rham et al.(2009)]{2009PhRvL.103p1601D} 
C.~de~Rham, J.~Khoury, and A.~Tolley, Phys.\ Rev.\ Lett.\ {\bf 103}, 161601(2009).

%28
\bibitem[de Rham et al.(2008)]{2008PhRvL.100y1603D} 
C.~de~Rham, G.~Dvali, S.~Hofmann, J.~Khoury, O.~Pujol{\`a}s, M.~Redi and A.~J.~Tolley, Phys.\ Rev.\ Lett.\ {\bf 100}, 251603 (2008).

%29\cite{Nicolis:2008in}
\bibitem{Nicolis:2008in}
  A.~Nicolis, R.~Rattazzi and E.~Trincherini,
  %``The Galileon as a local modification of gravity,''
  Phys.\ Rev.\  D {\bf 79}, 064036 (2009)
  [arXiv:0811.2197 [hep-th]].
  %%CITATION = PHRVA,D79,064036;%%

%30
\bibitem[Deffayet et al.(2009)]{2009PhRvD..79h4003D} 
C.~Deffayet, G.~Esposito-Far{\`e}se and A.~Vikman,
%Covariant Galileon
 \prd {\bf 79}, 084003 (2009), arXiv:0901.1314v2 [hep-th].

%31
\bibitem[Deffayet et al.(2009)]{2009PhRvD..80f4015D} 
C.~Deffayet, S.~Deser G.~Esposito-Far{\`e}se,
%Generalized Galileons: All scalar models whose curved background extensions maintain second-order field equations and stress tensors
 \prd {\bf 80}, 064015 (2009), arXiv:0906.1967v2 [gr-qc].

%32\cite{Chow:2009fm}
\bibitem{Chow:2009fm}
  N.~Chow and J.~Khoury,
  %``Galileon Cosmology,''
  Phys.\ Rev.\  D {\bf 80}, 024037 (2009)
  [arXiv:0905.1325 [hep-th]].
  %%CITATION = PHRVA,D80,024037;%%

%33
\bibitem[Silva
\& Koyama(2009)]{2009PhRvD..80l1301S} F.~P.~Silva and K.~Koyama,
%Self-accelerating universe in Galileon cosmology
 \prd {\bf 80}, 121301 (2009), arXiv:0909.4538v2 [astro-ph.CO].

%34
\bibitem[Deffayet et al.(2010)]{2010PhRvD..82f1501D} 
C.~Deffayet, S.~Deser and G.~Esposito-Far{\`e}se,
%Arbitrary p-form Galileons
 \prd {\bf 82}, 061501 (2010), arXiv:1007.5278v2 [gr-qc].

%35
\bibitem[Hinterbichler et al.(2010)]{2010PhRvD..82l4018H} 
K.~Hinterbichler, M.~Trodden and D.~Wesley,
%Multi-field galileons and higher co-dimension branes
 \prd {\bf 82}, 124018 (2010), arXiv:1008.1305v2 [hep-th].

%36
\bibitem[Padilla et al.(2010)]{2010JHEP...12..031P}
%Bi-galileon theory I: motivation and formulation
 A.~Padilla, P.~M.~Saffin and S.-Y.~Zhou, Journal of High Energy Physics {\bf 12}, 31 (2010), arXiv:1007.5424v4 [hep-th].

%37
\bibitem[Padilla et al.(2011)]{2011JHEP...01..099P} 
A.~Padilla, P.~M.~Saffin and  S.-Y.~Zhou,
%Bi-galileon theory II: phenomenology
 Journal of High Energy Physics {\bf 1}, 99 (2011), arXiv:1008.3312v4 [hep-th].

%38
\bibitem[Goon et al.(2011)]{2011PhRvD..83h5015G}  
G.~L.~Goon, K.~Hinterbichler and M.~Trodden,
%Stability and superluminality of spherical DBI galileon solutions
 \prd {\bf 83}, 085015 (2011), arXiv:1008.4580v2 [hep-th].

%39
\bibitem[Andrews et al.(2011)]{2011PhRvD..83d4042A} 
M.~Andrews, K.~Hinterbichler, J.~Khoury and M.~Trodden,
%Instabilities of Spherical Solutions with Multiple Galileons and SO(N) Symmetry
 \prd {\bf 83}, 044042 (2011), arXiv:1008.4128v2 [hep-th].

%40\cite{Khoury:2011da}
\bibitem{Khoury:2011da}
  J.~Khoury, J.~L.~Lehners and B.~A.~Ovrut,
  %``Supersymmetric Galileons,''
  arXiv:1103.0003 [hep-th].
  %%CITATION = ARXIV:1103.0003;%%

%41\cite{Trodden:2011xh}
\bibitem{Trodden:2011xh}
  M.~Trodden and K.~Hinterbichler,
  %``Generalizing Galileons,''
  arXiv:1104.2088 [hep-th].
  %%CITATION = ARXIV:1104.2088;%%

\bibitem{Woodard}
N.~C.~Tsamis and R.~P.~Woodard,
  %``Relaxing the cosmological constant,''
  Phys.\ Lett.\  B {\bf 301}, 351 (1993).
  %%CITATION = PHLTA,B301,351;%%
  
\bibitem{ABM}
V.~F.~Mukhanov, L.~R.~W.~Abramo and R.~H.~Brandenberger,
  %``On the Back reaction problem for gravitational perturbations,''
  Phys.\ Rev.\ Lett.\  {\bf 78}, 1624 (1997)
  [arXiv:gr-qc/9609026];\\
  %%CITATION = PRLTA,78,1624;%%
  L.~R.~W.~Abramo, R.~H.~Brandenberger and V.~F.~Mukhanov,
  %``The Energy - momentum tensor for cosmological perturbations,''
  Phys.\ Rev.\  D {\bf 56}, 3248 (1997)
  [arXiv:gr-qc/9704037].
  %%CITATION = PHRVA,D56,3248;%%
  
\bibitem{RHBrev2002}
R.~H.~Brandenberger,
  %``Back reaction of cosmological perturbations and the cosmological constant
  %problem,''
  arXiv:hep-th/0210165.
  %%CITATION = HEP-TH/0210165;%%
  
%42\cite{Brandenberger:1988aj}
\bibitem{Brandenberger:1988aj}
  R.~H.~Brandenberger, C.~Vafa,
  %``Superstrings in the Early Universe,''
  Nucl.\ Phys.\  {\bf B316}, 391 (1989).

%43\cite{Nayeri:2005ck}
\bibitem{Nayeri:2005ck}
  A.~Nayeri, R.~H.~Brandenberger and C.~Vafa,
  %``Producing a scale-invariant spectrum of perturbations in a Hagedorn phase
  %of string cosmology,''
  Phys.\ Rev.\ Lett.\  {\bf 97}, 021302 (2006)
  [arXiv:hep-th/0511140].
  %%CITATION = PRLTA,97,021302;%%
  
%44\cite{Brandenberger:2006xi}
\bibitem{Brandenberger:2006xi}
  R.~H.~Brandenberger, A.~Nayeri, S.~P.~Patil and C.~Vafa,
  %``Tensor Modes from a Primordial Hagedorn Phase of String Cosmology,''
  Phys.\ Rev.\ Lett.\  {\bf 98}, 231302 (2007)
  [arXiv:hep-th/0604126].
  %%CITATION = PRLTA,98,231302;%%
  
%45\cite{Brandenberger:2006pr}
\bibitem{Brandenberger:2006pr}
  R.~H.~Brandenberger {\it et al.},
  %``More on the spectrum of perturbations in string gas cosmology,''
  JCAP {\bf 0611}, 009 (2006)
  [arXiv:hep-th/0608186].
  %%CITATION = JCAPA,0611,009;%%
  
%46\cite{Battefeld:2005av}
\bibitem{Battefeld:2005av}
  T.~Battefeld and S.~Watson,
  %``String gas cosmology,''
  Rev.\ Mod.\ Phys.\  {\bf 78}, 435 (2006)
  [arXiv:hep-th/0510022].
  %%CITATION = RMPHA,78,435;%%

%47\cite{Brandenberger:2008nx}
\bibitem{Brandenberger:2008nx}
  R.~H.~Brandenberger,
  %``String Gas Cosmology,''
  [arXiv:0808.0746 [hep-th]].

%48\cite{Gasperini:1992em}
\bibitem{Gasperini:1992em}
  M.~Gasperini and G.~Veneziano,
  %``Pre - big bang in string cosmology,''
  Astropart.\ Phys.\  {\bf 1}, 317 (1993)
  [arXiv:hep-th/9211021].
  %%CITATION = APHYE,1,317;%%
  
%49\cite{Gasperini:2002bn}
\bibitem{Gasperini:2002bn}
  M.~Gasperini, G.~Veneziano,
  %``The Pre - big bang scenario in string cosmology,''
  Phys.\ Rept.\  {\bf 373}, 1-212 (2003).
  [hep-th/0207130].
  
%50\cite{Gasperini:2007vw}
\bibitem{Gasperini:2007vw}
  M.~Gasperini, G.~Veneziano,
  %``String Theory and Pre-big bang Cosmology,''
  [hep-th/0703055].

%51\cite{Khoury:2001wf}
\bibitem{Khoury:2001wf}
  J.~Khoury, B.~A.~Ovrut, P.~J.~Steinhardt and N.~Turok,
  %``The Ekpyrotic universe: Colliding branes and the origin of the hot big
  %bang,''
  Phys.\ Rev.\  D {\bf 64}, 123522 (2001)
  [arXiv:hep-th/0103239].
  %%CITATION = PHRVA,D64,123522;%%
  
%52\cite{Khoury:2001bz}
\bibitem{Khoury:2001bz}
  J.~Khoury, B.~A.~Ovrut, N.~Seiberg, P.~J.~Steinhardt and N.~Turok,
  %``From big crunch to big bang,''
  Phys.\ Rev.\  D {\bf 65}, 086007 (2002)
  [arXiv:hep-th/0108187].
  %%CITATION = PHRVA,D65,086007;%%
  
%53\cite{Brandenberger:2001bs}
\bibitem{Brandenberger:2001bs}
  R.~Brandenberger and F.~Finelli,
  %``On the spectrum of fluctuations in an effective field theory of the
  %Ekpyrotic universe,''
  JHEP {\bf 0111}, 056 (2001)
  [arXiv:hep-th/0109004].
  %%CITATION = JHEPA,0111,056;%%

%54\cite{Steinhardt:2001st}
\bibitem{Steinhardt:2001st}
  P.~J.~Steinhardt and N.~Turok,
  %``Cosmic evolution in a cyclic universe,''
  Phys.\ Rev.\  D {\bf 65}, 126003 (2002)
  [arXiv:hep-th/0111098].
  %%CITATION = PHRVA,D65,126003;%%
  
%55\cite{Finelli:2002we}
\bibitem{Finelli:2002we}
  F.~Finelli,
  %``Assisted contraction,''
  Phys.\ Lett.\  B {\bf 545}, 1 (2002)
  [arXiv:hep-th/0206112].
  %%CITATION = PHLTA,B545,1;%%

%56\cite{Tsujikawa:2002qc}
\bibitem{Tsujikawa:2002qc}
  S.~Tsujikawa, R.~Brandenberger and F.~Finelli,
  %``On the construction of nonsingular pre - big bang and ekpyrotic cosmologies
  %and the resulting density perturbations,''
  Phys.\ Rev.\  D {\bf 66}, 083513 (2002)
  [arXiv:hep-th/0207228].
  %%CITATION = PHRVA,D66,083513;%%

%57\cite{Tolley:2003nx}
\bibitem{Tolley:2003nx}
  A.~J.~Tolley, N.~Turok and P.~J.~Steinhardt,
  %``Cosmological perturbations in a big crunch / big bang space-time,''
  Phys.\ Rev.\  D {\bf 69}, 106005 (2004)
  [arXiv:hep-th/0306109].
  %%CITATION = PHRVA,D69,106005;%%  

%58\cite{Khoury:2003rt}
\bibitem{Khoury:2003rt}
  J.~Khoury, P.~J.~Steinhardt and N.~Turok,
  %``Designing cyclic universe models,''
  Phys.\ Rev.\ Lett.\  {\bf 92}, 031302 (2004)
  [arXiv:hep-th/0307132].
  %%CITATION = PRLTA,92,031302;%%

%59\cite{Khoury:2004xi}
\bibitem{Khoury:2004xi}
  J.~Khoury,
  %``A Briefing on the ekpyrotic / cyclic universe,''
  arXiv:astro-ph/0401579.
  %%CITATION = ASTRO-PH/0401579;%%

%60\cite{Creminelli:2004jg}
\bibitem{Creminelli:2004jg}
  P.~Creminelli, A.~Nicolis and M.~Zaldarriaga,
  %``Perturbations in bouncing cosmologies: Dynamical attractor versus scale
  %invariance,''
  Phys.\ Rev.\  D {\bf 71}, 063505 (2005)
  [arXiv:hep-th/0411270].
  %%CITATION = PHRVA,D71,063505;%%

\bibitem{Novello}
M.~Novello and S.~E.~P.~Bergliaffa,
  %``Bouncing Cosmologies,''
  Phys.\ Rept.\  {\bf 463}, 127 (2008)
  [arXiv:0802.1634 [astro-ph]].
  %%CITATION = PRPLC,463,127;%%
  
\bibitem{RHBrev2011}
R.~H.~Brandenberger,
  %``Introduction to Early Universe Cosmology,''
  PoS {\bf ICFI2010}, 001 (2010)
  [arXiv:1103.2271 [astro-ph.CO]].
  %%CITATION = POSCI,ICFI2010,001;%%
  
%61\cite{Steinhardt:2008nk}
\bibitem{Steinhardt:2008nk}
  P.~J.~Steinhardt and D.~Wesley,
  %``Dark Energy, Inflation and Extra Dimensions,''
  Phys.\ Rev.\  D {\bf 79}, 104026 (2009)
  [arXiv:0811.1614 [hep-th]].
  %%CITATION = PHRVA,D79,104026;%%

%62\cite{ArkaniHamed:2003uy}
\bibitem{ArkaniHamed:2003uy}
  N.~Arkani-Hamed, H.~C.~Cheng, M.~A.~Luty and S.~Mukohyama,
  %``Ghost condensation and a consistent infrared modification of gravity,''
  JHEP {\bf 0405}, 074 (2004)
  [arXiv:hep-th/0312099].
  %%CITATION = JHEPA,0405,074;%%

%63\cite{Dubovsky:2005xd}
\bibitem{Dubovsky:2005xd}
  S.~Dubovsky, T.~Gregoire, A.~Nicolis and R.~Rattazzi,
  %``Null energy condition and superluminal propagation,''
  JHEP {\bf 0603}, 025 (2006)
  [arXiv:hep-th/0512260].
  %%CITATION = JHEPA,0603,025;%%

%64\cite{ArkaniHamed:2003uz}
\bibitem{ArkaniHamed:2003uz}
  N.~Arkani-Hamed, P.~Creminelli, S.~Mukohyama and M.~Zaldarriaga,
  %``Ghost inflation,''
  JCAP {\bf 0404}, 001 (2004)
  [arXiv:hep-th/0312100].
  %%CITATION = JCAPA,0404,001;%%

%65\cite{Creminelli:2006xe}
\bibitem{Creminelli:2006xe}
  P.~Creminelli, M.~A.~Luty, A.~Nicolis and L.~Senatore,
  %``Starting the Universe: Stable Violation of the Null Energy Condition and
  %Non-standard Cosmologies,''
  JHEP {\bf 0612}, 080 (2006)
  [arXiv:hep-th/0606090].
  %%CITATION = JHEPA,0612,080;%%

%66\cite{Buchbinder:2007ad}
\bibitem{Buchbinder:2007ad}
  E.~I.~Buchbinder, J.~Khoury and B.~A.~Ovrut,
  %``New Ekpyrotic cosmology,''
  Phys.\ Rev.\  D {\bf 76}, 123503 (2007)
  [arXiv:hep-th/0702154].
  %%CITATION = PHRVA,D76,123503;%%

%67\cite{Lin:2010pf}
\bibitem{Lin:2010pf}
  C.~Lin, R.~H.~Brandenberger and L.~P.~Levasseur,
  %``A Matter Bounce By Means of Ghost Condensation,''
  JCAP {\bf 1104}, 019 (2011)
  [arXiv:1007.2654 [hep-th]].
  %%CITATION = JCAPA,1104,019;%%

%68\cite{Creminelli:2010ba}
\bibitem{CNT}
  P.~Creminelli, A.~Nicolis and E.~Trincherini,
  %``Galilean Genesis: An Alternative to inflation,''
  JCAP {\bf 1011}, 021 (2010)
  [arXiv:1007.0027 [hep-th]].
  %%CITATION = JCAPA,1011,021;%%

\bibitem{Ellis}
G.~F.~R.~Ellis and R.~Maartens,
  %``The emergent universe: Inflationary cosmology with no singularity,''
  Class.\ Quant.\ Grav.\  {\bf 21}, 223 (2004)
  [arXiv:gr-qc/0211082].
  %%CITATION = CQGRD,21,223;%%
  
%69
\bibitem{Durrer:1994}
  R.~Durrer and M.~Sakellariadou,
  %A new Contribution to Cosmological Perturbations of some Inflationary Models
   Phys.\ Rev.\  D {\bf 50}, 6115-6122 (2007), [arXiv:astro-ph/9404043v1].

\bibitem{Guth}
A. Guth, 
%  ``The Inflationary Universe: A Possible Solution To The Horizon And Flatness
% Problems,''
  Phys.\ Rev.\  D {\bf 23}, 347 (1981).
  %%CITATION = PHRVA,D23,347;%%

\bibitem{Reheatrev}
R.~Allahverdi, R.~Brandenberger, F.~Y.~Cyr-Racine and A.~Mazumdar,
  %``Reheating in Inflationary Cosmology: Theory and Applications,''
  Ann.\ Rev.\ Nucl.\ Part.\ Sci.\  {\bf 60}, 27 (2010)
  [arXiv:1001.2600 [hep-th]].
  
  %%CITATION = ARNUA,60,27;%%
\bibitem{Dolgov}
 A. Dolgov and A. Linde,
 % ``Baryon Asymmetry In Inflationary Universe,''
  Phys.\ Lett.\  B {\bf 116}, 329 (1982);
  %%CITATION = PHLTA,B116,329;%%
  
\bibitem{AFW}
 L. Abbott,  E. Farhi and M. Wise. 
 % ``Particle Production In The New Inflationary Cosmology,''
  Phys.\ Lett.\  B {\bf 117}, 29 (1982);
  %%CITATION = PHLTA,B117,29;%%
  
\bibitem{TB}
J. Traschen and R. Brandenberger,
%  ``Particle Production During Out-Of-Equilibrium Phase Transitions,''
  Phys.\ Rev.\  D {\bf 42}, 2491 (1990).
  %%CITATION = PHRVA,D42,2491;%%

\bibitem{DK}
A. Dolgov and D. Kirilova,
%  ``Production of particles by a variable scalar field,''
  Sov.\ J.\ Nucl.\ Phys.\  {\bf 51}, 172 (1990)
  [Yad.\ Fiz.\  {\bf 51}, 273 (1990)].
  %%CITATION = YAFIA,51,273;%%
  
\bibitem{KLS1}
L. Kofman, A. Linde  and A. Starobinsky,
%  ``Reheating after inflation,''
  Phys.\ Rev.\ Lett.\  {\bf 73}, 3195 (1994).
  %%CITATION = PRLTA,73,3195;%%

\bibitem{STB}
Y. Shtanov, J. Traschen and R. Brandenberger,
%  ``Universe reheating after inflation,''
  Phys.\ Rev.\  D {\bf 51}, 5438 (1995).
  %%CITATION = PHRVA,D51,5438;%%

\bibitem{KLS2}
L. Kofman, A. Linde and A. Starobinsky,
%  ``Towards the theory of reheating after inflation,''
  Phys.\ Rev.\  D {\bf 56}, 3258 (1997).
  %%CITATION = PHRVA,D56,3258;%%

\bibitem{MFB}
V.~F.~Mukhanov, H.~A.~Feldman and R.~H.~Brandenberger,
%  ``Theory of cosmological perturbations. Part 1. Classical perturbations. Part
%  2. Quantum theory of perturbations. Part 3. Extensions,''
  Phys.\ Rept.\  {\bf 215}, 203 (1992).
  %%CITATION = PRPLC,215,203;%%
  
\bibitem{RHBrev2004}
R.~H.~Brandenberger,
%  ``Lectures on the theory of cosmological perturbations,''
  Lect.\ Notes Phys.\  {\bf 646}, 127 (2004)
  [arXiv:hep-th/0306071].
  %%CITATION = LNPHA,646,127;%%
 
 \bibitem{Martineau}
 P.~Martineau,
%  ``On the decoherence of primordial fluctuations during inflation,''
  Class.\ Quant.\ Grav.\  {\bf 24}, 5817 (2007)
  [arXiv:astro-ph/0601134].
  %%CITATION = CQGRD,24,5817;%% 

 \bibitem{Kiefer}    
C.~Kiefer, I.~Lohmar, D.~Polarski and A.~A.~Starobinsky,
%  ``Pointer states for primordial fluctuations in inflationary cosmology,''
  Class.\ Quant.\ Grav.\  {\bf 24}, 1699 (2007)
  [arXiv:astro-ph/0610700].
  %%CITATION = CQGRD,24,1699;%%
  
 \bibitem{Grishchuk}
L.~P.~Grishchuk,
%  ``Amplification Of Gravitational Waves In An Istropic Universe,''
  Sov.\ Phys.\ JETP {\bf 40}, 409 (1975)
  [Zh.\ Eksp.\ Teor.\ Fiz.\  {\bf 67}, 825 (1974)].
  %%CITATION = ZETFA,67,825;%% 
  
\bibitem{Galileoninflation}
T.~Kobayashi, M.~Yamaguchi, J.~Yokoyama,
  %``G-inflation: Inflation driven by the Galileon field,''
  Phys.\ Rev.\ Lett.\  {\bf 105}, 231302 (2010).
  [arXiv:1008.0603 [hep-th]] ;\\
 C.~Burrage, C.~de Rham, D.~Seery, A.~J.~Tolley,
  %``Galileon inflation,''
  JCAP {\bf 1101}, 014 (2011).
  [arXiv:1009.2497 [hep-th]].

\end{thebibliography}
\end{document}